\documentclass[preprint]{aastex}

\received{}
\accepted{}

\slugcomment{13 May 2003 Revision}

\author{
Gordon T. Richards,\altaffilmark{1,2}
Patrick B. Hall,\altaffilmark{2,3}
Daniel E. Vanden Berk,\altaffilmark{4}
Michael A. Strauss,\altaffilmark{2}
Donald P. Schneider,\altaffilmark{1}
Michael A. Weinstein,\altaffilmark{1}
Timothy A. Reichard,\altaffilmark{1}
Donald G. York,\altaffilmark{5,6}
G.R. Knapp,\altaffilmark{2}
Xiaohui Fan,\altaffilmark{7}
\v{Z}eljko Ivezi\'{c},\altaffilmark{2}
J. Brinkmann,\altaffilmark{8}
Tam\'as Budav\'ari,\altaffilmark{9,10}
Istv\'an Csabai,\altaffilmark{9,10}
and R. C. Nichol\altaffilmark{11}
}

\altaffiltext{1}{Department of Astronomy and Astrophysics, The Pennsylvania State University, 525 Davey Laboratory, University Park, PA 16802.}
\altaffiltext{2}{Princeton University Observatory, Peyton Hall, Princeton, NJ 08544.}
\altaffiltext{3}{Departamento de Astronom\'{\i}a y Astrof\'{\i}sica, Pontificia Universidad Cat\'{o}lica de Chile, Casilla 306, Santiago 22, Chile.}
\altaffiltext{4}{Department of Physics and Astronomy, University of Pittsburgh, 3941 O'Hara Street, Pittsburgh, PA 15260.}
\altaffiltext{5}{Department of Astronomy and Astrophysics, The University of Chicago, 5640 South Ellis Avenue, Chicago, IL 60637.}
\altaffiltext{6}{Enrico Fermi Institute, The University of Chicago, 5640 South Ellis Avenue, Chicago, IL 60637.}
\altaffiltext{7}{Steward Observatory, University of Arizona, 933 North Cherry Avenue, Tucson, AZ 85721.}
\altaffiltext{8}{Apache Point Observatory, P.O. Box 59, Sunspot, NM 88349.}
\altaffiltext{9}{Department of Physics of Complex Systems, E\"otv\"os University, Budapest, Pf.\ 32, H-1518 Budapest, Hungary.}
\altaffiltext{10}{Department of Physics and Astronomy, The Johns Hopkins University, 3400 North Charles Street, Baltimore, MD 21218-2686.}
\altaffiltext{11}{Department of Physics, Carnegie Mellon University, 5000 Forbes Avenue, Pittsburgh, PA 15232.}

\begin{document}

\title{Red and Reddened Quasars in the Sloan Digital Sky Survey}

\begin{abstract}

We investigate the overall continuum and emission line properties of
quasars as a function of their optical/UV spectral energy
distributions.  Our sample consists of 4576 quasars from the Sloan
Digital Sky Survey (SDSS) that were chosen using homogeneous selection
criteria.  Expanding on our previous work which demonstrated that the
optical/UV color distribution of quasars is roughly Gaussian, but with
a red tail, here we distinguish between 1) quasars that have
intrinsically blue (optically flat) power-law continua, 2) quasars
that have intrinsically red (optically steep) power-law continua, and
3) quasars whose colors are inconsistent with a single power-law
continuum.  We find that 273 (6.0\%) of the quasars in our sample fall
into the latter category and appear to be redder because of SMC-like
dust extinction and reddening rather than because of synchrotron
emission.  Even though the SDSS Quasar Survey is optically selected
and flux-limited, we demonstrate that it is sensitive to dust reddened
quasars with $E(B-V)\lesssim0.5$ assuming a classical Small Magellanic
Cloud (SMC) extinction curve.  The color distribution of our SDSS
quasar sample suggests that the population of moderately dust
reddened, but otherwise normal (i.e., Type 1) quasars is smaller than
the population of unobscured quasars: we estimate that a further 10\%
of the quasar population with $M_i<-25.61$ is missing from the SDSS
sample because of extinction, bringing the total fraction of dust
reddened quasars to 15\% of broad-line quasars.

We also investigate the emission and absorption line properties of
these quasars as a function of color and comment on how some of these
results relate to Boroson \& Green type eigenvectors.  Quasars with
intrinsically red (optically steep) power-law continua tend to have
narrower Balmer lines and weaker \ion{C}{4}, \ion{C}{3}], \ion{He}{2}
and $3000\,{\rm \AA}$ bump emission as compared with bluer (optically
flatter) quasars.  The change in strength of the $3000\,{\rm \AA}$
bump appears to be dominated by the Balmer continuum and not by
\ion{Fe}{2} emission.  The dust reddened quasars have even narrower
Balmer lines and weaker $3000\,{\rm \AA}$ bumps, in addition to having
considerably larger equivalent widths of [\ion{O}{2}] and [\ion{O}{3}]
emission.  The fraction of broad absorption line quasars (BALQSOs)
increases from $\sim 3.4\%$ for the bluest quasars to perhaps as large
as $20\%$ for the dust reddened quasars, but the intrinsic color
distribution will be much bluer if all BALQSOs are affected by dust
reddening.

\end{abstract}


\section{Introduction}

The existence of a significant population of red quasars has been the
subject of debate for a number of years.  The color distribution of
quasars \markcite{wfp+95,fww00,btb+01,rfs+01}({Webster} {et~al.} 1995; {Francis}, {Whiting}, \& {Webster} 2000; {Brotherton} {et~al.} 2001; {Richards} {et~al.} 2001) and the discovery of a few
examples of highly reddened radio-selected quasars
\markcite{glw+02,whb+03}({Gregg} {et~al.} 2002; {White et al.} 2003) strongly suggest that a sizeable population of
such objects does indeed exist.  Nevertheless, it is still unclear how
often different mechanisms (e.g., dust extinction, a synchrotron
emission turnover, or simply intrinsic continuum differences) are at
work, whether this class of objects is a significant fraction of the
total quasar population, and whether moderately dust reddened Type 1
quasars are significant contributors to the X-ray background
\markcite{mcb+00,bhs+00}({Mushotzky} {et~al.} 2000; {Brandt} {et~al.} 2000) --- independent of the contribution of Type 2
quasars with completely obscured broad emission line regions (BELRs).

Part of the reason that these issues are not better understood may be
the heterogeneous definitions that have been used to classify quasars
as ``red''.  This paper suggests that traditional samples of red
quasars that impose cuts on observed colors or spectral index
\markcite{wfp+95,fdw+01,wsc+02,glw+02}(e.g., {Webster} {et~al.} 1995; {Francis} {et~al.} 2001; {Wilkes} {et~al.} 2002; {Gregg} {et~al.} 2002) are too heterogeneous
with redshift to yield physically homogeneous samples of red quasars.
Spectral indices are sensitive to the choice of continuum windows, and
simple color cuts necessarily ignore the significant contribution that
emission and absorption lines make to the colors of quasars as a
function of redshift.  That is not to say that such red quasar
criteria are not reasonable, but rather that care must be taken in the
interpretation of such samples after the fact.  Herein we address
these issues with a sample of quasars from the Sloan Digital Sky
Survey (SDSS; \markcite{yaa+00}{York} {et~al.} 2000).  Using this sample, we present a more
redshift-independent selection criterion for red quasars that allows
for better classification of the types (and causes) of red quasars,
which, in turn, will lead to a better census of the quasar population
in the future.

We demonstrate that the distribution of optical colors of normal
quasars is resolved by the SDSS and we isolate quasars as a function
of optical color.  The majority of our quasars are consistent with a
roughly Gaussian distribution in color that can be characterized by a
Gauassian distribution in spectral index.  For these quasars, which
have colors that are roughly consistent with a single power-law
continuum, the fact that our photometric errors are smaller than the
width of the distribution allows us to differentiate between those
quasars that have bluer than average colors (hereafter referred to as
``optically flat'' quasars) and quasars that have slightly redder than
average color --- but still blue in an absolute sense --- hereafter
referred to as ``optically steep'' quasars.  In addition, there is a
red tail to the distribution beyond that which can be explained by a
Gaussian distribution in spectral index.  These quasars that have
continua with curvature that cannot be explained by a single power-law
slope; they appear reddened as a result of a decrement of blue flux
such as might result from dust extinction and are referred to
hereafter as ``dust reddened'' quasars.  Quasars can also appear
redder than average as a result of an excess of red flux such as might
result from the presence of a synchrotron source with a high-frequency
turnover in the rest-frame UV or optical, such quasars are referred to
hereafter as ``red synchrotron'' quasars; \markcite{sr97,fww00}{Serjeant} \& {Rawlings} 1997; {Francis} {et~al.} 2000).  See
\S~\ref{sec:restframesel} and \S~\ref{sec:compspec} for figures that
help with visual clarification of our flat/steep/dusty nomenclature.

Our analysis of reddened quasars discusses whether the heavily
reddened quasars in the literature are just the tail of a population
dominated by less reddened quasars, or if there is an equal
distribution of extreme, moderate, and mild reddening, or if the
distribution is bimodal with reddening.  We demonstrate that the SDSS
quasar survey is sensitive to moderately dust reddened quasars ($0.1
\lesssim E(B-V) \lesssim 0.5$) and that this sensitivity allows us to
comment on the population of even more heavily reddened, but otherwise
normal Type 1 quasars.  We cannot, however, comment on the extent of
the heavily obscured Type 2 quasar population using this data set.

Further analysis of the color distribution of quasars and the
relationship of the emission and absorption line properties to
optical/UV color may prove useful in determining the underlying
physics of AGN.  For example, it has become popular to probe the
physical characteristics of quasars such as the black hole mass,
luminosity, and accretion rate through eigenvector-type analyses
\markcite{bg92,wlb+99,smz+02}({Boroson} \& {Green} 1992; {Wills} {et~al.} 1999; {Sulentic} {et~al.} 2002).  Such analyses attempt to relate different
quasar observables to each other and to determine which physical
mechanisms might cause such correlations.  In this paper, we
specifically examine the relationship of optical/UV color to some of
those emission and absorption line properties that are used in
eigenvector analysis, especially since optical/UV color has not been
one of the primary properties used in such analyses \markcite{bf99}({Brotherton} \& {Francis} 1999).

In \S~2 we describe the dataset used herein and discuss the ability of
the SDSS to find red quasars.  Section~3 presents the method by which
we divide quasars into color samples and also presents our definition
of an apparently dust reddened quasar.  Section~4 presents and
examines composite quasar spectra as a function of color.  A
discussion is presented in \S~5, and conclusions are given in \S~6.

\section{The Data}

\subsection{Sample Definition}

The quasars were selected for SDSS spectroscopic followup from the
SDSS imaging survey which uses a wide-field multi-CCD camera
\markcite{gcr+98}({Gunn} {et~al.} 1998); quasar candidates were identified using the SDSS
Quasar Target Selection Algorithm \markcite{rfn+02}({Richards} {et~al.} 2002a).  Objects were
selected based on their broad-band SDSS colors
\markcite{fig+96,lgs99,hfs+01,stk+02,slb+02}({Fukugita} {et~al.} 1996; {Lupton}, {Gunn}, \& {Szalay} 1999; {Hogg} {et~al.} 2001; {Smith} {et~al.} 2002; {Stoughton} {et~al.} 2002) and as optical matches to
radio-detected quasars from the VLA ``FIRST'' survey \markcite{bwh95}({Becker}, {White}, \& {Helfand} 1995),
although the objects are primarily radio quiet \markcite{imk+02}({Ivezi{\' c}} {et~al.} 2002).  The
automated extraction of point-spread-function magnitudes from the
images is discussed by \markcite{lgi+01}{Lupton} {et~al.} (2001), and \markcite{pmh+02}{Pier} {et~al.} (2003) present the
details of the SDSS astrometric solution.  Magnitudes (and therefore
colors) in this paper have been corrected for Galactic reddening using
the reddening maps from \markcite{sfd98}{Schlegel}, {Finkbeiner}, \&  {Davis} (1998).  The quasar spectra cover
$3800\,{\rm \AA}$ to $9200\,{\rm \AA}$ at a resolution of $1800$ to
$2000$.  The spectra are calibrated to spectrophotometric standard
F-stars and are thus corrected for Galactic reddening (at the scale of
the separation to the nearest calibration star, $\lesssim1\deg$) in
the limit that these F stars are behind the dust in the Galaxy.  The
spectrophotometric calibration includes scaling by wider aperture
``smear'' exposures to the photometry from the imaging data.  See
\markcite{slb+02}{Stoughton} {et~al.} (2002) for more details regarding the spectroscopic system and
analysis pipelines.  A discussion of the spectroscopic tiling
algorithm can be found in \markcite{blm+02}{Blanton et al.} (2002).

The quasars used in this paper are the first SDSS quasars to be
selected in a homogeneous manner using the algorithm described in
detail by \markcite{rfn+02}{Richards} {et~al.} (2002a); most SDSS quasars to date were selected with
an older version of the algorithm (see \markcite{slb+02}{Stoughton} {et~al.} 2002).  This sample
is magnitude limited to $i^*=19.1$\footnote{The data used herein use
preliminary photometric calibrations, thus we follow the convention of
\markcite{slb+02}{Stoughton} {et~al.} (2002) and use, for example, $i^*$, to refer to measured
magnitudes and $i$ to refer to the filters themselves.}.  We
supplement our main sample by including quasars that were observed as
part of the SDSS Southern Survey (which is designed to go deeper than
the Main Survey) during the fall of 2001.  These quasars were also
selected with the algorithm described by \markcite{rfn+02}{Richards} {et~al.} (2002a) except that we
relaxed some of the constraints.  Quasar candidates were selected to
$i^*=19.9$ (for both color and radio selection) and we did not bias
against some high contamination regions of color space (e.g., where
white dwarfs are found).  Between the two samples, there are 5803
quasars with $M_{i^*}<-23$ (${\rm H_o = 50\,km\,s^{-1}\,Mpc^{-1}},
\Omega = 1, \Lambda = 0$, $\alpha_{\nu}=-0.5$
[$f_\nu\propto\nu^{\alpha_{\nu}}$]) that had at least one emission
line broader than $1000\,{\rm km\,s^{-1}}$ FWHM.  These quasars come
from 147 spectroscopic plates (which are 3 degrees in diameter and
hold 640 spectroscopic fibers) covering an area of $\sim600\,{\rm
deg}^2$.  To define a sample that is as homogeneous as possible and
that is not contaminated by light from the host galaxy or the absence
of flux in the Ly$\alpha$ forest, we will further restrict the sample
to $0.3 \le z \le 2.2$, leaving 4576 quasars in our sample (639 of
which are fainter than $i^*=19.1$).

\subsection{The SDSS's Sensitivity to Red/Reddened Quasars \label{sec:sdssred}}

Many previous (and current) optical surveys of quasars are somewhat
biased against red quasars --- and especially dust reddened quasars
--- because they are magnitude-limited at the blue end of the optical
spectrum (e.g., the $B$ band at $4400\,{\rm \AA}$).  The SDSS, on the
other hand, is magnitude limited in the $i$ band ($\lambda_{\rm eff}
\approx 7460\,{\rm \AA}$ for a quasar with optical spectral index
$\alpha_{\nu}=-0.5$).  As a result, the SDSS is more sensitive to dust
reddened quasars since their spectral shapes are a strong function of
wavelength.  In addition, the SDSS photometric errors are much smaller
than the errors of the photographic photometry used for other
wide-area optical surveys of quasars.  These smaller errors mean that
there is less scatter in the stellar locus and thus less of quasar
parameter space is contaminated by stars.  Second, to the extent
allowed by a limited number of spectroscopic fibers, we consider
objects outside of the stellar locus (in color space) to be quasar
candidates, thus we are sensitive to quasars with ``nonstandard''
colors; see \markcite{rfn+02}{Richards} {et~al.} (2002a) for details.  Finally, the wide wavelength
coverage and high quality of the SDSS spectra allow us to identify
quasars with unusual spectral energy distributions.

Figure~11 of \markcite{rfn+02}{Richards} {et~al.} (2002a) showed the completeness of the SDSS quasar
target selection algorithm as a function of redshift and absolute
magnitude for 53,000 simulated quasars \markcite{fan99}({Fan} 1999) with a Gaussian
distribution of spectral indices, $\alpha_{\nu}=-0.5\pm0.3$.  We test
the SDSS's sensitivity to optically steep quasars and dust reddened
quasars by modifying the colors of this initial sample.  To create a
test sample of optically steep quasars, we have simply shifted the
median of the power law spectral index of the 53,000 simulated quasar
colors to $\alpha_{\nu}=-1.5$.  To simulate dust reddened quasars, we
reddened and extincted the colors using a SMC-like dust reddening law
\markcite{plp+84}({Prevot} {et~al.} 1984) with the form $R_{\lambda} \equiv
A_{\lambda}/E(B-V)=1.39\lambda^{-1.2}$, where $\lambda$ is in microns
and the dust is located at the redshift of the quasar.

\begin{figure}[p]
\epsscale{1.0} 
\plottwo{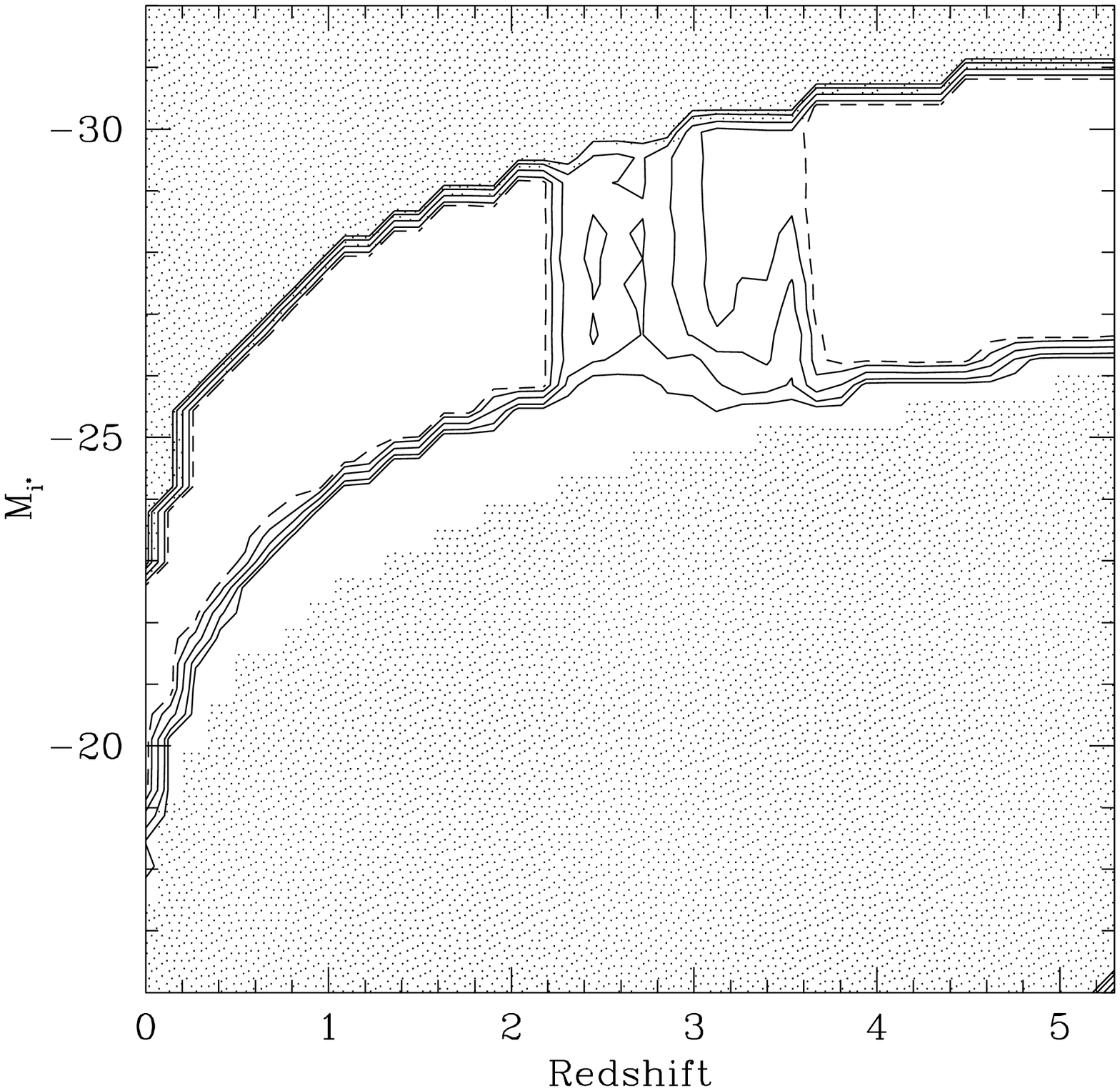}{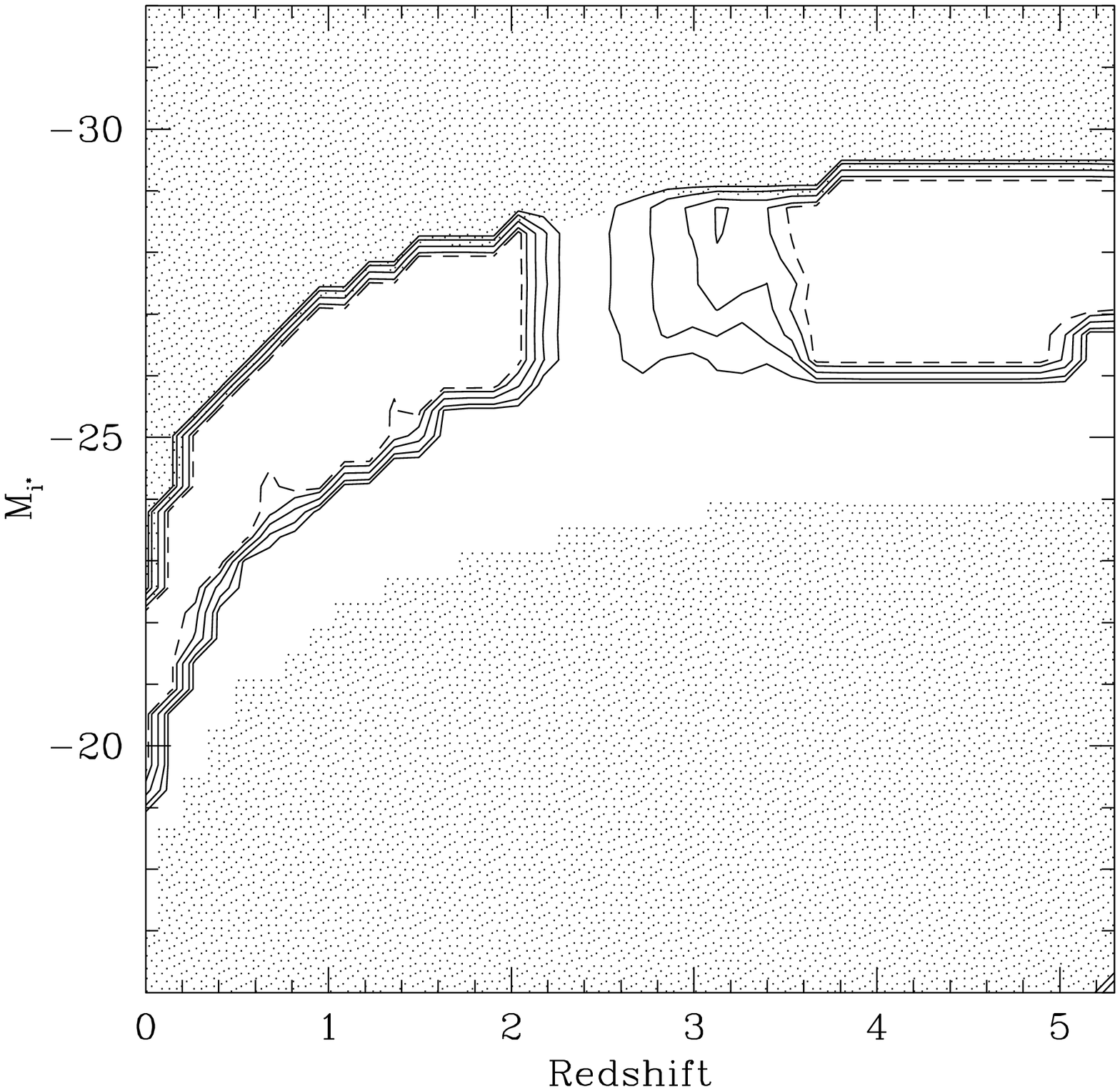}
\caption{{\em Left:} Completeness for simulated red (optically steep)
but unextincted quasars with $\alpha_{\nu}=-1.5\pm0.3$. {\em Right:}
Completeness for simulated dust reddened and extincted quasars with
$E(B-V)=0.1$ (assuming SMC dust extinction).  In both panels, contours
are drawn at 10, 25, 50, 75 and 90\% completeness, with the 90\%
completeness contour shown as a dashed line. The shaded region is
where there are no objects in the simulations; quasars more luminous
than the most luminous of our simulated quasars are clearly even more
likely to be detected. \label{fig:fig1}}
\end{figure}

The results of running these simulated optically steep and dust
reddened quasars through the SDSS's quasar target selection algorithm
\markcite{rfn+02}({Richards} {et~al.} 2002a) are shown in Figure~\ref{fig:fig1}.  The contours show
the likelihood of selecting a quasar at a given redshift and
$M_{i^*}$; the shaded regions are parts of the parameter space that
are not explored by these simulations.  We find that even optically
steep quasars with $\alpha_{\nu}=-1.5$ and moderately dust reddened
quasars with $E(B-V)=0.1$ (characteristic of our dust reddened quasar
criterion, see \S~\ref{sec:arq}) have high selection probabilities in
the SDSS.  For $\alpha_{\nu}=-1.5$, the detection probability is over
90\% for the entire redshift range considered in this paper
($0.3<z<2.2$).  Even for $E(B-V)=0.1$, the detection probability is at
least 90\% for the $z\lesssim2$ quasars studied herein and also for
$z\gtrsim3.5$ quasars.  The sensitivity of the SDSS for high-redshift
quasars that are redder than average is particularly good since their
red colors move them {\em away} from the colors of stars.

Under certain circumstances, the SDSS is also sensitive to more
heavily dust reddened quasars.  The dynamic range of the SDSS quasar
target selection algorithm is approximately 4 magnitudes
($15.0<i^*<19.1$) for $z\lesssim3$, so at $z=2$, we are sensitive to
intrinsically luminous quasars with $E(B-V)$ at least as large as
$0.54$.  In most cases, such heavily reddened quasars will be
extincted out of our sample, but they may still be discovered by the
SDSS if their unreddened magnitude is $i^*<15$ (the SDSS bright limit
for spectroscopy) and dust causes them to be extincted {\em into} our
sample.  Quasars as intrinsically bright as $i^*<15$ are rare, but
there will be enough of them in the area of sky covered by the SDSS
that we can use them to comment on the extent of heavily reddened, but
otherwise normal, broad-lined quasars.

Two examples probe the ability of the SDSS to detect extremely
reddened quasars that are extincted by $E(B-V)\sim0.5$.  PMN
0134$-$0931 ($z=2.2$, $i^*=19.65$) is a heavily reddened,
$E(B-V)\sim0.54$, gravitationally lensed quasar \markcite{glw+02,wmh+02}({Gregg} {et~al.} 2002; {Winn} {et~al.} 2002)
recovered by the $griz$ branch of the SDSS color selection algorithm;
see \markcite{hry+02}{Hall} {et~al.} (2002b).  Although this quasar is magnified by lensing, it
still demonstrates the ability of the SDSS to find intrinsically
luminous, dust reddened quasars.  Perhaps more noteworthy in the
context of this paper is the SDSS's recovery of FIRST~J0738+2750
\markcite{glw+02}({Gregg} {et~al.} 2002), an unlensed $z=1.985$ quasar that is internally
reddened by $E(B-V) \sim 0.5$.  Both quasars are fainter than the
magnitude limit of the Main Survey, but are bright enough to be
included in the deeper extension of the spectroscopic survey in the
southern equatorial region.  Even more heavily reddened and/or
obscured quasars, including broad absorption line quasars, such as
those presented by \markcite{hal+02}{Hall} {et~al.} (2002a), may not always be classified as
quasars by the automated SDSS pipelines as a result of their unusual
spectra; see \S~\ref{sec:arq}.

\section{Defining Quasar Samples as a Function of Color}

\subsection{Observed-Color Selection and Redshift Heterogeneity}

\begin{figure}[p]
\epsscale{1.0}
\plotone{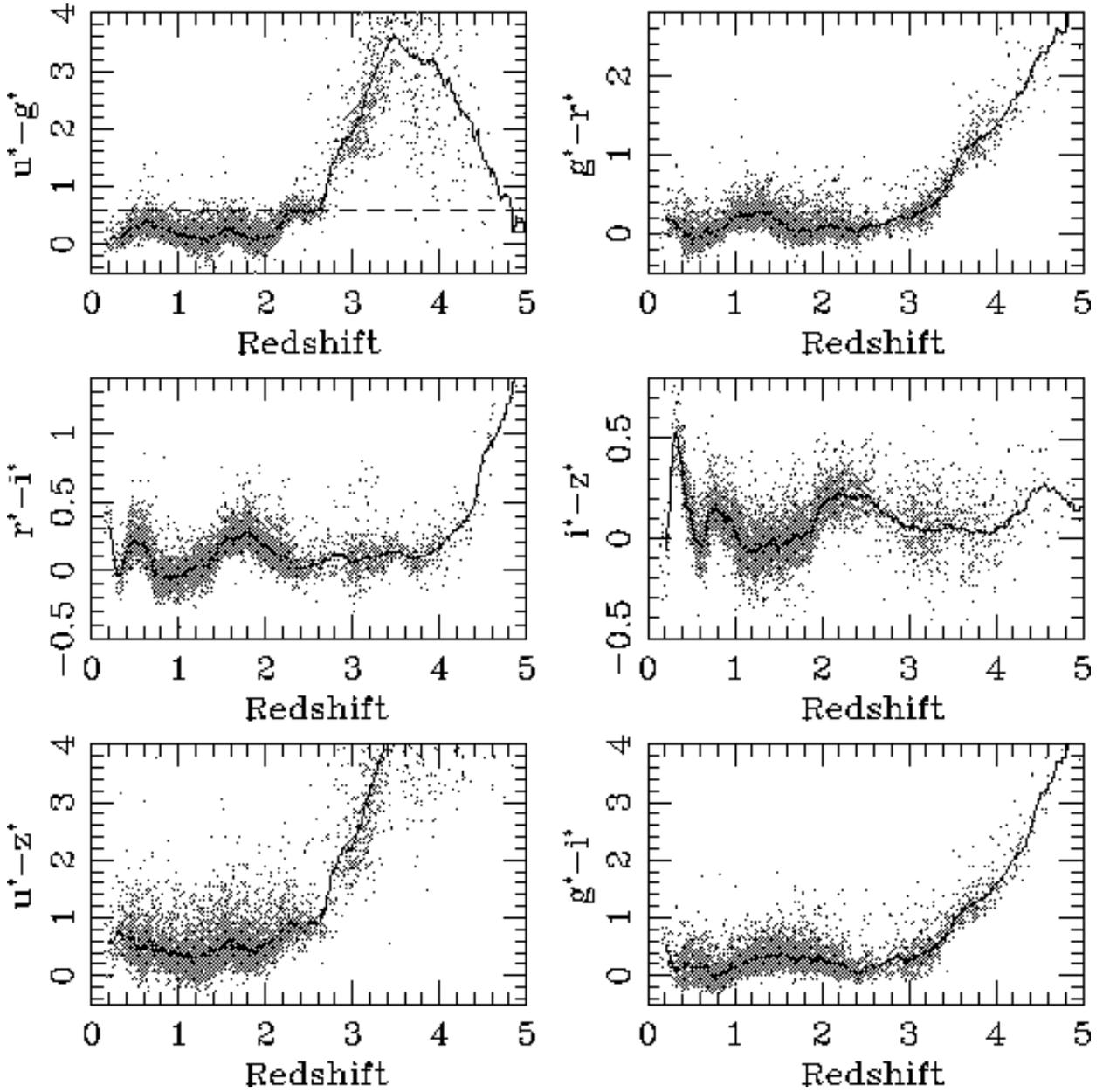}
\caption{Colors of all 5803 SDSS quasars in our initial sample ({\em
grey points}).  The black lines show the median colors as a function
of redshift.  The dashed black line in the upper left hand panel is at
$u^*-g^* = 0.6$, which might be used as a dividing line between
``red'' and normal quasars.  Note how the use of such a criterion
produces a heterogeneous sample as a result of color changes as a
function of redshift produced by emission and absorption features.  A
quasar with $u^*-g^* = 0.6$ can be perfectly normal (if somewhat
optically steep) at $z=0.6$, dust reddened at $z=1.2.$, or appear red
because of Ly$\alpha$ forest absorption at
$z\gtrsim3$. \label{fig:fig2}}
\end{figure}

Red quasars are typically searched for using a simple color cut in the
observed frame, such as $B-K\gtrsim5$.  In the upper left hand panel
of Figure~\ref{fig:fig2} we illustrate such an observed frame color
cut (specfically, $u^*-g^*=0.6$).  Although such a criterion does
indeed select quasars that are {\em apparently} red, it fails to
distinguish between four possible cases: (1) that the quasar has an
intrinsically red power-law continuum (i.e., is intrinsically
optically steep), (2) that the quasar is reddened by dust extinction,
(3) that the quasar has excess synchrotron emission, and (4) that the
redshift of the quasar is such that the quasar appears to be redder
than normal because of strong emission or absorption in one of the
bands (e.g., Ly$\alpha$ emission, absorption by the Ly$\alpha$ forest,
or BAL troughs).  A further example is that of Figure~5 of
\markcite{bh01}{Barkhouse} \& {Hall} (2001) in which they demonstrate that while using $J-K>2.0$ as a
color-cut may well select dust reddened quasars, it also selects
perfectly normal, if somewhat optically steep, quasars with $z<0.5$.
Thus it is perhaps not unexpected that follow-up X-ray observations of
such a sample of ``red'' quasars \markcite{wsc+02}({Wilkes} {et~al.} 2002) finds heterogeneous
X-ray properties among red quasars selected in this manner.

There are similar problems with defining red quasars according to
spectral indices.  For example \markcite{glw+02}{Gregg} {et~al.} (2002), suggest $\alpha<-1$
($f_{\nu} \propto \nu^{\alpha}$) as an appropriate definition of a red
quasar.  The problem with this approach is that most spectra do not
have enough wavelength coverage for redshift-independent spectral
index determination.  \markcite{vrb+01}{Vanden Berk} {et~al.} (2001) found that there were good
continuum windows only at $\sim1450\,{\rm \AA}$ and $\sim4000\,{\rm
\AA}$; most other commonly used windows are contaminated by
\ion{Fe}{2} emission.  These wavelength regions are only seen
simultaneously at $1.30 \leq z \leq 1.62$ even in the SDSS spectra,
which have broader wavelength coverage than do most other optical
quasar studies.  Furthermore, the optical spectral index of quasars is
a function of the continuum windows used, and therefore depends on
redshift \markcite{vrb+01,srf+02,prp+03}(see, e.g., {Vanden Berk} {et~al.} 2001; {Schneider} {et~al.} 2002; {Pentericci et al.} 2003).  Finally, using
the spectral index can fail to distinguish between quasars that are
intrinsically optically steep and quasars that are dust reddened.
Dust reddened quasars will have spectral curvature that a simple
power-law fit will be unable to characterize.

Observed-color selection is clearly necessary when initiating a
follow-up campaign when the redshifts are not yet known.  However,
interpretation of results from red quasar samples requires the
construction of good statistical samples that are independent of
redshift.  Thus, it is clear that it would be very useful to have a
definition for a ``red'' quasar that takes redshift into account.  In
particular, we desire a simple but effective selection criterion that
depends only on the SDSS photometry (given a redshift), yet correlates
strongly with the overall optical/UV spectral energy distribution
(SED) of quasars independent of the effect of emission lines.  We will
show that statistically significant differences in the optical/UV
continua of quasars are reflected in their {\em relative colors} (the
colors referenced to the median color at a given redshift; see below).
We also demonstrate these relative colors can be used to distinguish
between intrinsically optically steep quasars and dust reddened
quasars.

\subsection{Rest Frame Selection Using Relative Colors \label{sec:restframesel}}

\begin{figure}[p]
\epsscale{1.0}
\plotone{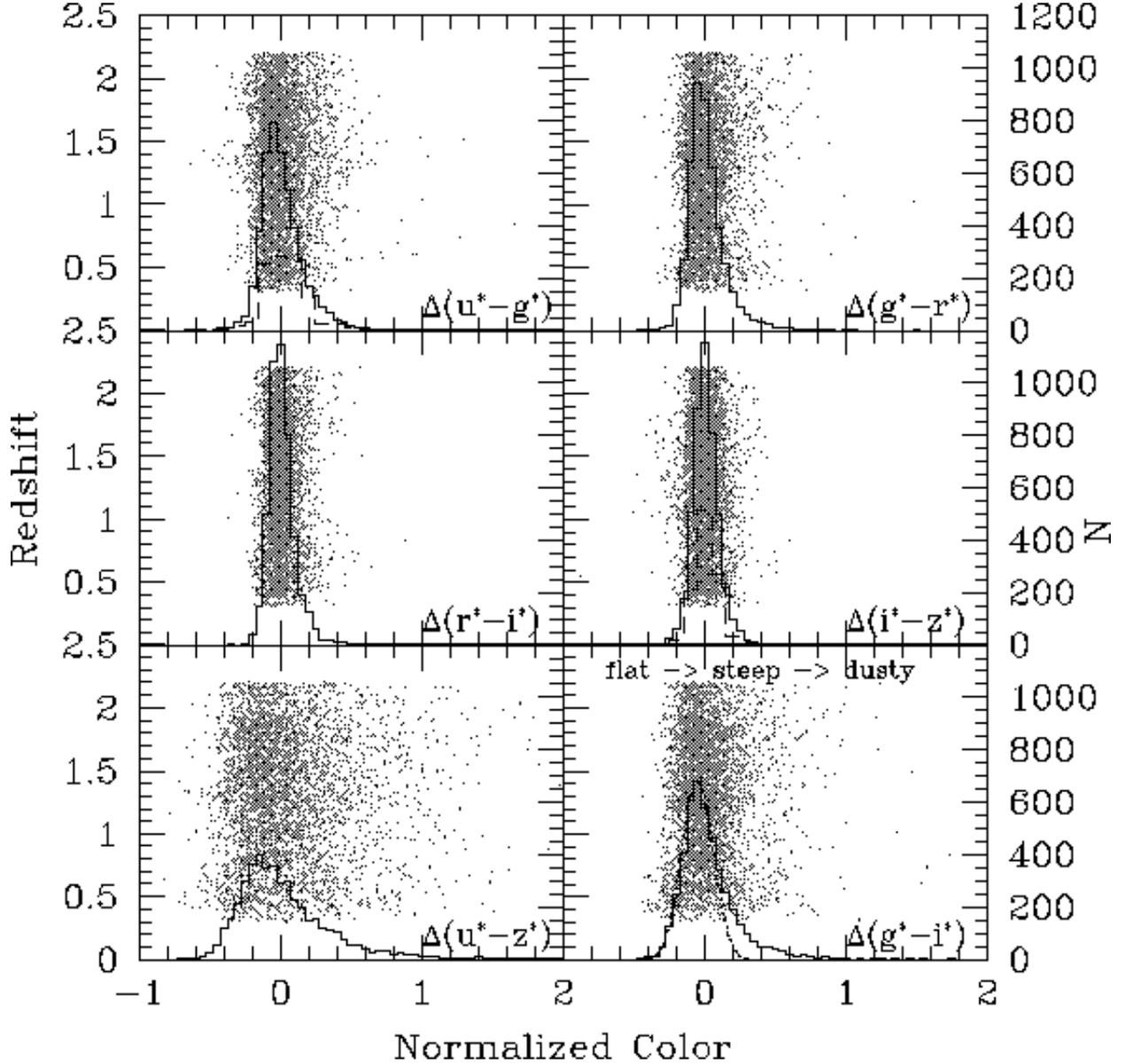}
\caption{Relative colors of the 4576 quasars with $0.3 \le z \le 2.2$,
$i^*<19.9$, and $M_{i^*}<-23.0$.  The grey points are a scatter plot
of the relative colors ($x$-axes) versus the redshift (left hand
$y$-axes).  The black histogram is a histogram of the number of
quasars as a function of relative color; the number in each bin is
given by the right-hand $y$-axes.  The dashed histograms in the
$\Delta (u^*-g^*)$ and $\Delta (i^*-z^*)$ panels show the relative
color distribution of radio-detected quasars (binned at half the
resolution and multiplied by a factor of 5).  The dotted line in the
lower right hand panel shows a Gaussian color distribution as fit to
the peak and blue wing; note the clear excess of quasars to the red.
The top of the $\Delta (g^*-i^*)$ panel shows the progression of
optically-flat to optically-steep to dust reddened in our adopted
nomenclature.
\label{fig:fig3}}
\end{figure}

Rather than determining the continuum color of our quasars by
measuring a spectral index from each spectrum, we determine the
underlying continuum color by subtracting the median colors of quasars
at the redshift of each quasar from the measured colors of each
quasar; we will refer to this difference as a {\em relative color}
\markcite{rfs+01}({Richards} {et~al.} 2001).\footnote{We choose not to call this difference a
``color excess'', since color excesses are defined to be positive,
whereas relative colors can be either positive or negative; see
Figure~\ref{fig:fig3}.  Note that previously, in \markcite{rfs+01}{Richards} {et~al.} (2001), we
referred to it as the normalized color.} Figure~\ref{fig:fig2} shows
the measured and median colors as a function of redshift in bins of
$\Delta z = 0.01$ for the SDSS quasars in our sample.  As shown by
\markcite{rfs+01}{Richards} {et~al.} (2001), the features in the color-redshift diagrams can be
understood as being due to emission lines moving in and out of the
filters.  For redshifts between $z=0.6$ and $z=2.2$, we expect the
colors of quasars to be dominated by their power law continua and
emission lines (as opposed to light from the host galaxy for $z<0.6$
and Ly$\alpha$ forest absorption for $z>2.2$).  We also chose to
include quasars with $0.3 < z < 0.6$ so that we sampled both H$\alpha$
and H$\beta$ (in an attempt to study the Balmer decrement); however,
in the end, the H$\alpha$ line was too poorly sampled to measure
Balmer decrements.

Figure~\ref{fig:fig3} depicts the distributions of the relative
colors.  The gray points give the individual relative colors as a
function of redshift; see \markcite{rfs+01}{Richards} {et~al.} (2001) for similar plots as a
function of apparent and absolute magnitude.  There are no trends in
$\Delta (g^*-i^*)$ with position on the sky or \markcite{sfd98}{Schlegel} {et~al.} (1998) Galactic
reddening.  Furthermore, these color distributions are not
predominantly due to photometric errors; the distributions are
formally {\em resolved}.  For example, the FWHM of the
$\Delta(r^*-i^*)$ distribution is $\sim0.2$ and even at $i^*\sim19$
the photometric error in $r^*-i^*$ is typically only $\sim0.025$ (FWHM
$\sim0.06$).  As a result, we can meaningfully distinguish between
blue (optically flat) and red (optically steep) quasars.

As we shall discuss below, from Figure~\ref{fig:fig3} it is quite
clear that there is a tail of red quasars which is particularly
pronounced in colors involving the $u$ or $g$ filters.  (This tail was
previously seen in a smaller SDSS sample by \markcite{rfs+01}{Richards} {et~al.} 2001.)  This
tail indicates that the scatter in the colors of quasars is not a
simple Gaussian distribution around the median color as can be seen in
the lower right hand panel of Figure~\ref{fig:fig3} where we show that
a Gaussian distribution ({\em dotted line}) can be made to match the
blue extent of the distribution, but not the red.

\subsection{Photometric Spectral Indices}

To determine the overall optical/UV SEDs of our quasars, we fit a
power-law to the five SDSS photometric data points in a manner similar
to that described by \markcite{wwf01}{Whiting}, {Webster}, \& {Francis} (2001).  Instead of using the observed
photometric data points, which will deviate from the underlying
power-law continuum because of emission and absorption features, we
make use of the relative colors.  First, we normalize all of the
quasars to have the same apparent $i$ band magnitude since we are only
interested in the slope and shape of the continuum.  Then we compute
the other {\em relative magnitudes} from the relative colors.  In
essence, we have applied a $k$-correction for the emission lines.  To
determine the spectral index from the relative magnitudes, $m_{rel}$,
we utilized the error-weighted linear least-squares fitting routine
from \markcite{ptv+92}{Press} {et~al.} (1992) and solved for what we refer to as the {\em
photometric spectral index}, $\alpha_p$, according to
\begin{equation}
-0.4 \times m_{rel} = \alpha_p \log \lambda + C.
\end{equation}
using the original magnitude errors as measured in each band and
effective wavelengths of 3541, 4653, 6147, 7461, and $8904\,{\rm \AA}$
in $ugriz$ (appropriate for a power-law continuum with
$\alpha_{\lambda}=-1.5$ and $f_{\lambda}\propto\lambda^{\alpha}$).  By
subtracting the median quasar color from all the observed colors, we
have essentially {\em defined} the relative magnitudes such that the
median quasar will now have $f_{\lambda} \propto \lambda^{-2}$ (i.e.,
$\alpha_{\nu}\equiv0$).

\begin{figure}[p]
\epsscale{1.0}
\plottwo{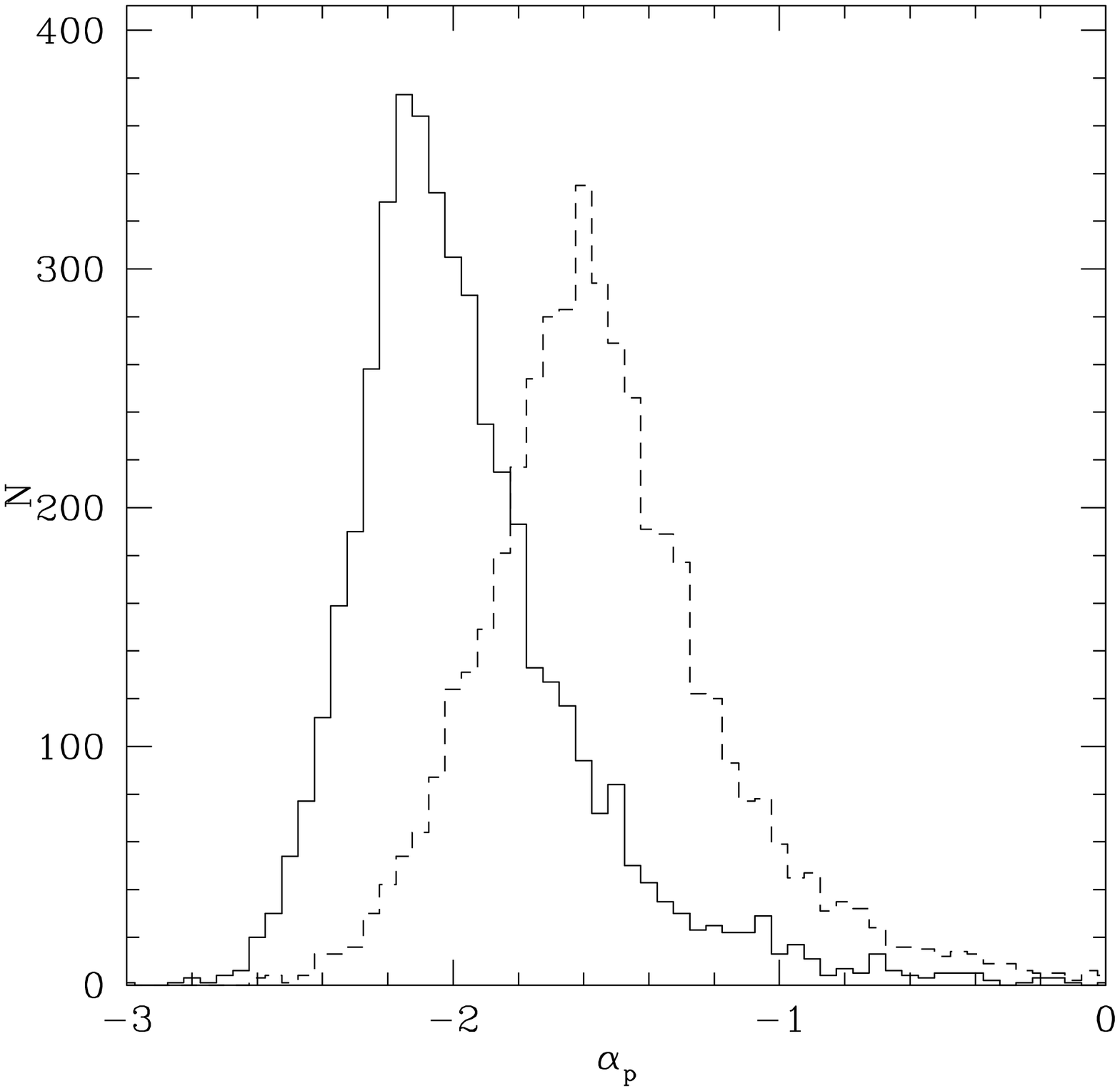}{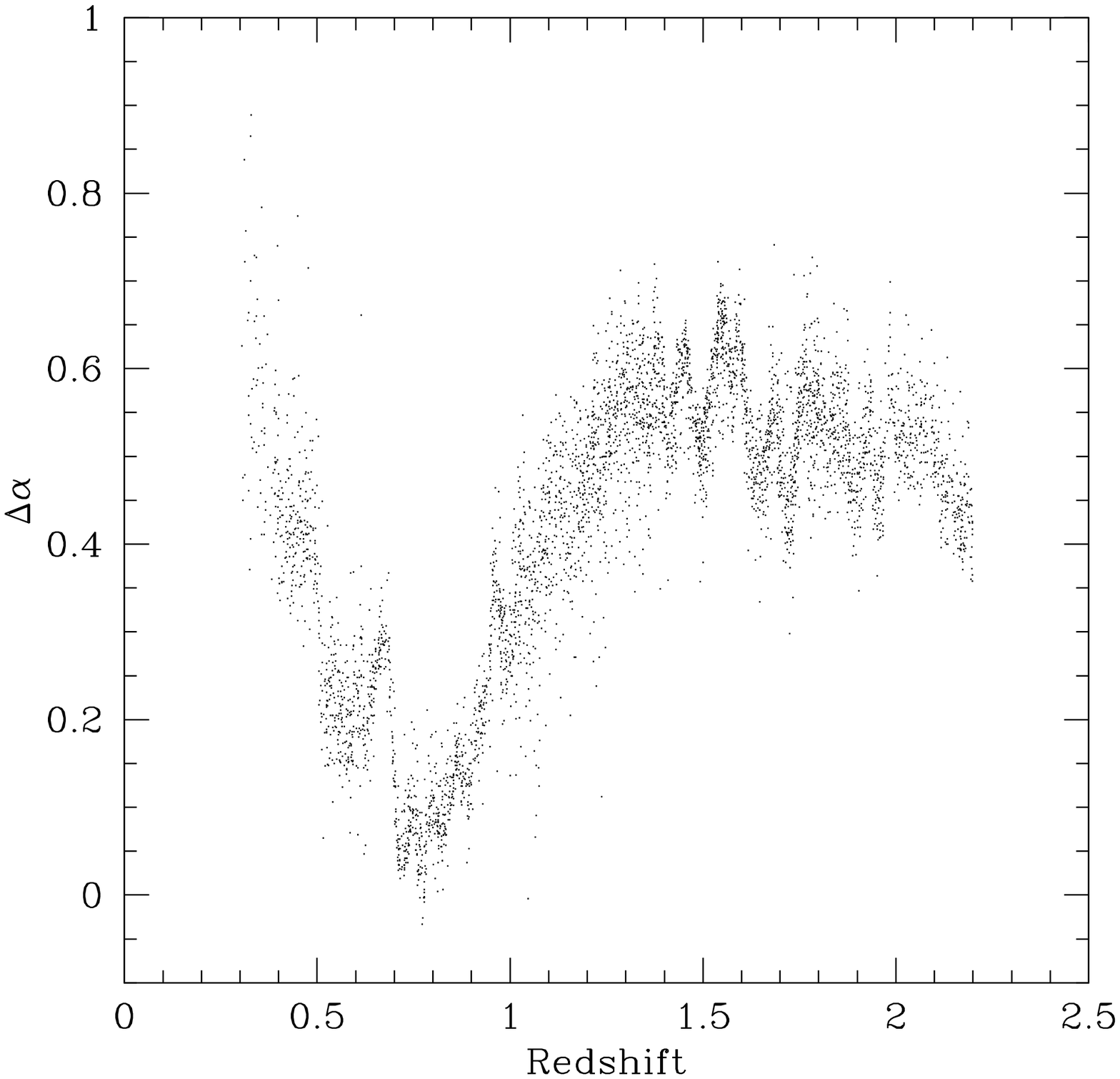}
\caption{{\em Left:} Histogram of photometric spectral indices
computed from the relative magnitudes of each quasar (solid line).
The dashed line shows the histogram of spectral indices computed using
the observed instead of relative magnitudes. {\em Right:} Difference
between redshift corrected photometric spectral index and uncorrected
photometric spectral index as a function of redshift.  The expected
difference is $\sim0.5$; larger values indicate that the uncorrected
photometric spectral index is too red.\label{fig:fig4}}
\end{figure}

The solid line in the left panel of Figure~\ref{fig:fig4} shows the
resulting distribution of photometric spectral indices, while the
dashed line shows the spectral index distributions derived from
fitting a power-law to the {\em observed} magnitudes in a manner
similar to \markcite{wwf01}{Whiting} {et~al.} (2001).  The mean offset between the distributions
is irrelevant; it merely arises because the relative colors are
defined to have a median of $\alpha_{p}=-2$.  The right panel of
Figure~\ref{fig:fig4} shows the difference between photometric
spectral indices using relative magnitudes and observed magnitudes as
a function of redshift.  Even though both indices recover the tail of
red quasars, it is clear that failure to correct for redshift
dependent color effects can lead to erroneous photometric spectral
indices, where the errors result from not accounting for emission and
absorption features (especially the small blue bump) and are
systematic with redshift.

\subsection{Using Relative Colors To Define Color Samples \label{sec:arq}} 

\begin{figure}[p]
\epsscale{1.0}
\plotone{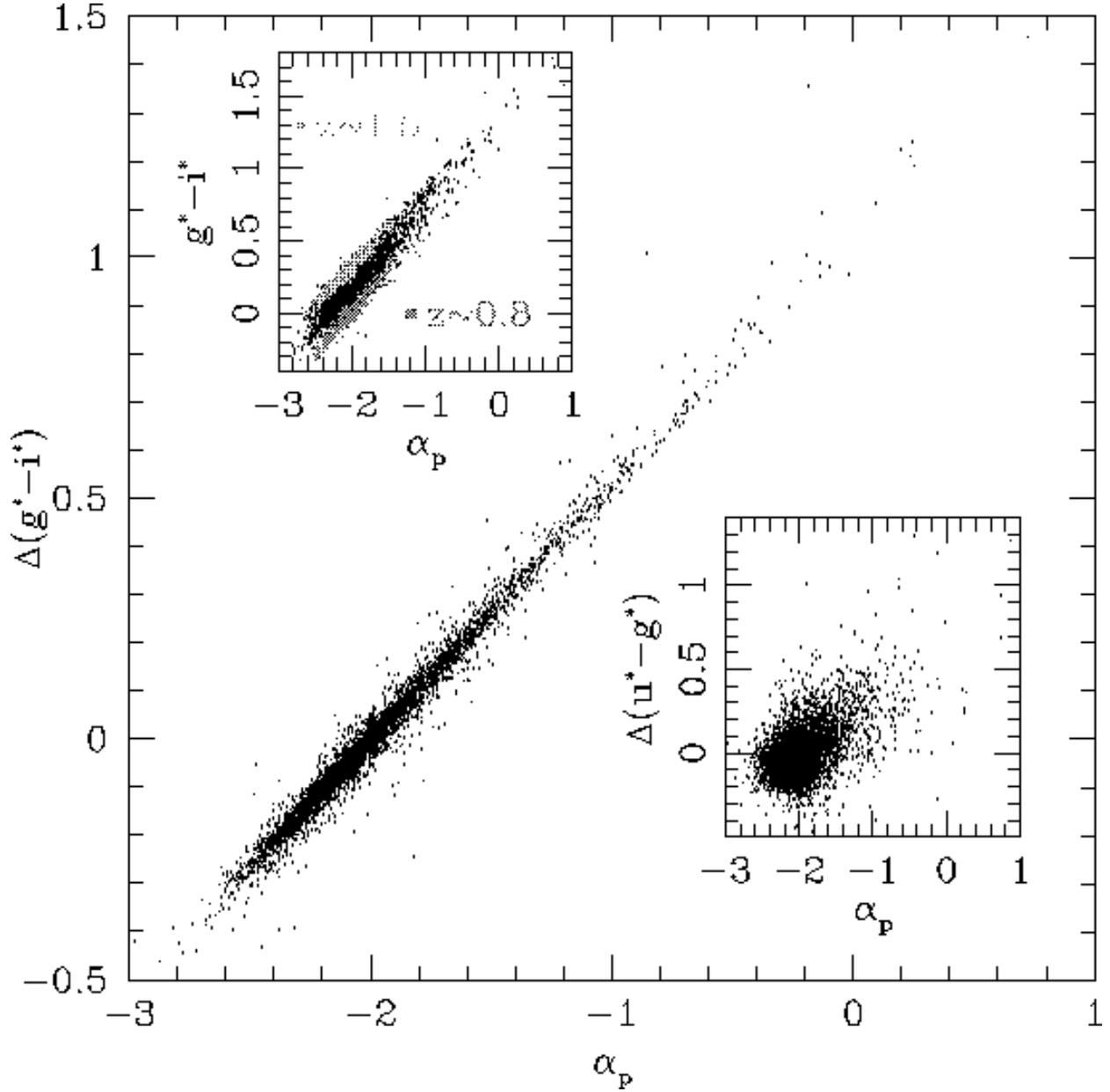}
\caption{Relative $g^*-i^*$ colors versus photometric spectral
index, $\alpha_p$.  {\em Upper Left Inset:} Observed $g^*-i^*$ colors
versus $\alpha_p$.  The correlation between $g^*-i^*$ and $\alpha_p$
is not nearly as tight as for $\Delta(g^*-i^*)$ because of redshift
effects.  {\em Lower Right Inset:} Relative $u^*-g^*$ colors versus
$\alpha_p$, which shows that the relative $u^*-g^*$ colors are not a
good surrogate for $\alpha_p$.  \label{fig:fig5}}
\end{figure}

Having determined the broad-band SEDs of our quasars, we now ask if we
can find a cut (or cuts) using relative colors that can serve as a
surrogate for the procedure of finding the photometric spectral index.
Surprisingly, the relative $u^*-g^*$ color, $\Delta (u^*-g^*)$, is
not a very good discriminator despite the large spread in that
distribution (see the lower right inset of Fig.~\ref{fig:fig5}).  The
observed colors also do not serve as an adequate surrogate for
spectral index.  The upper left-hand panel of Figure~\ref{fig:fig5}
shows that observed $g^*-i^*$ indeed is correlated with spectral
index, but is broadened by redshift effects.  For example, a $z=1.5$
quasar with $g^*-i^*=0.2$ predicts a much bluer (flatter) spectral
index than a $z=0.8$ quasar with the same observed color.  On the
other hand, both the relative colors $\Delta (g^*-r^*)$ and $\Delta
(r^*-i^*)$ correlate well with the photometric spectral index and have
much less scatter than the observed colors.  The combination of the
two, $\Delta (g^*-i^*)$, is an excellent redshift-independent
surrogate for the photometric spectral index at $z<2.2$ as can be seen
in the main panel of Figure~\ref{fig:fig5}.

The quantity $\Delta (g^*-i^*)$ can also be used to distinguish
between quasars in the red tail and optically steep quasars (at least
in a statistical sense).  Figure~\ref{fig:fig3} shows that the
distribution of relative $r^*-i^*$ and $i^*-z^*$ colors is roughly
Gaussian and symmetric.  If the distribution of colors were purely the
result of a roughly Gaussian distribution of power-law spectral
indices, the distributions would be similarly Gaussian and symmetric
for the other relative colors.  However, as we have already mentioned,
the other colors, including $\Delta (g^*-i^*)$, show a distinctive
asymmetric tail to the red (as highlighted by excess above that of a
Gaussian distribution [{\em dotted line}] in the lower right hand
panel).  The objects in this tail are not consistent with an optically
steep power-law continuum.  This behavior {\em is} characteristic of
dust reddening, which gives rise to a curved SED.

Based upon the above discussion, we group our sample of quasars into
color classes based upon the relative $g^*-i^*$ color.  We begin by
isolating the quasars in the red tail of the color distribution, which
we presume to be reddened by dust.  The definition of such quasars
must address the fact that although power-laws are redshift invariant,
dust reddened power-laws are not (at least for typical dust extinction
curves).  Dust-reddened quasars at higher redshift will have redder
colors since a given set of filters will sample shorter rest
wavelengths, which are more affected by reddening.\footnote{Grey
extinction curves are theoretically possible but have never been
conclusively observed.}  Therefore, since we believe dust reddening is
largely responsible for this population, we must use redshift
dependent criteria to select them consistently.

\begin{figure}[p]
\epsscale{1.0}
\plotone{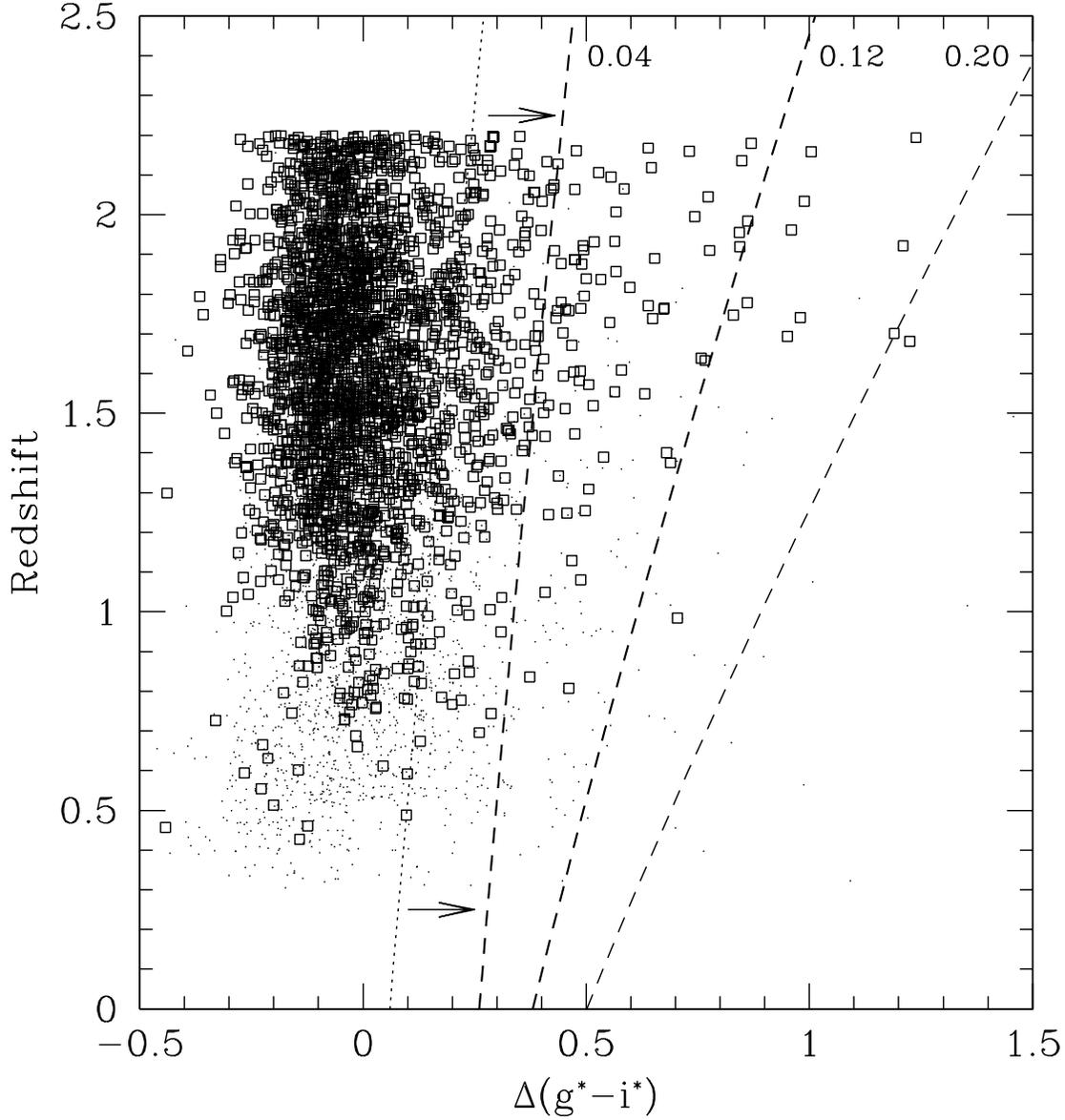}
\caption{Redshift versus relative $g^*-i^*$ color for quasars with
$M_{i^*}<-25.61$ ({\em open squares}) and $M_{i^*}\ge-25.61$ ({\em
small points}). The dotted line shows the effect of SMC-type reddening
as a function of redshift with $E(B-V)=0.04$.  The dashed lines are
for $E(B-V)=0.04$, $0.12$, and $0.2$, respectively, but are shifted
redward by 0.2 to match our dust reddened quasar definition.  The two
thicker dashed lines at $E(B-V)=0.04,0.12$ outline the part of the
dust reddened quasar region that we use to create our primary dust
reddened composite spectrum.
\label{fig:fig6}}
\end{figure}

Figure~\ref{fig:fig3} shows no trend of $\Delta(g^*-i^*)$ with
redshift.  Part of the reason why no trend is seen is that reddened
quasars with higher redshifts (which sample bluer rest wavelengths)
can be extincted enough that they drop out of the sample, whereas
lower redshift quasars with similarly large reddenings will not.  If
we impose a cut in absolute magnitude ($M_{i^{*}}<-25.61$) such that
the least luminous quasars could be seen at {\em all} redshifts in our
sample, we {\em do} see such a redshift trend.  This trend is shown in
Figure~\ref{fig:fig6}, where the $\Delta(g^*-i^*)$ colors of less
luminous quasars are shown by points and more luminous quasars by the
open squares.  The dashed lines show the expected change in color as a
function of redshift for an SMC-type reddening law
\markcite{plp+84,pei92}({Prevot} {et~al.} 1984; {Pei} 1992) with $E(B-V)=0.04,0.12,0.20$ (from left to right,
and shifted to the red by $0.20$ magnitudes; see caption and next
paragraph).  In our volume-limited sample ({\em open squares}),
quasars at higher redshifts are redder, as expected in a dust
reddening scenario.  There are simply very few low-redshift quasars in
our sample area which are intrinsically luminous enough to remain in
our sample once they are extincted by an amount corresponding to a
reddening of $\Delta(g^*-i^*)\gtrsim0.5$ at $z\lesssim1.5$.

As a result of this redshift dependence, we define a dust reddened
quasar to be any quasar that is redder than $\Delta(g^*-i^*)=0.2$ by
the redshift dependent correction for $E(B-V)=0.04$ (using the SMC
reddening law given above), see Table~1.  That is, all our putative
dust reddened quasars have colors redder than a quasar whose intrinsic
power-law continuum slope gives it a color of $\Delta(g^*-i^*)=0.2$
and that is also dust reddened by $E(B-V)=0.04$.  Our choice of
$E(B-V)=0.04$ is somewhat arbitrary, but it is effective in defining a
sample of quasars with color distinct from the general population.  In
practice, this yields a minimum value of $\Delta(g^*-i^*)\sim0.3$ for
the lowest redshift quasars in our dust reddened sample (see
Fig.~\ref{fig:fig6}).

For the sake of our analysis, we construct a restricted sample of dust
reddened quasars by imposing an upper limit to how red they can be
(see Table~1) since we are most sensitive to redder quasars at lower
redshifts and we would like our criterion to be consistent as a
function of redshift.  We exclude objects that are redder than
$\Delta(g^*-i^*)=0.2$ by more than $E(B-V)=0.12$ (as a function of
redshift) since our completeness drops rapidly for even more heavily
reddened quasars.  This restriction means that our restricted dust
reddened sample excludes the most heavily reddened and absorbed
quasars.  We will also examine a subsample that includes all of these
quasars, but even this sample will not include quasars of the type
discussed in \markcite{hal+02}{Hall} {et~al.} (2002a) since those quasars were all identified by
eye and not by the automated algorithm on which we have relied for our
sample.  Such quasars are also of significant interest and will be
discussed in future papers.  Of the 4576 quasars in our sample, 273
(6.0$\pm$0.4\%) are redder than the $E(B-V)=0.04$ cut and can be
considered to be dust reddened\footnote{If we restrict our sample to
those areas of the sky where the selection limit was $i\le19.1$ the
fraction drops to 4.5$\pm$0.4\% (170/3810).  Thus the dust reddened
fraction among the $i>19.1$ quasars is higher than 6\%, as might be
expected.}; 211 (4.6$\pm$0.3\%) also meet the $E(B-V)=0.12$ upper
limit condition that we impose for the sake of uniformity and will
constitute our restricted dust reddened sample in the following
sections.  (The uncertainties on the percentages above are statistical
only, and neglect the effects of systematic errors.)

Our redshift dependent definition of a dust reddened quasar
technically introduces a bias into the sample in that we have defined
the cut with the explicit assumption of dust reddening.  Clearly it
would be better to define dust reddened quasars in terms of their
rest-frame colors, but we cannot do so without interpolating (and
extrapolating) our data.  It could be argued that our definition
weakens our hypothesis that the sample is indeed dust reddened.  For
example, the non-LTE accretion disk models of \markcite{hab+00}{Hubeny} {et~al.} (2000) predict
quasi-thermal spectra with significant intrinsic curvature in the
rest-frame UV; such objects could fall within our dust reddened
definition.  In their model, lower luminosity objects also tend to
have greater curvature (their Figure~13).  This could explain why the
$M_{i^*}\ge-25.61$ objects in our Figure~\ref{fig:fig6} extend to
redder colors at $z\lesssim1.5$ than do the $M_{i^*}\le-25.61$
objects.  However, their model does not explain why the
$M_{i^*}\le-25.61$ objects extend to redder colors at $z\gtrsim1.5$
than at $z\lesssim1.5$.  Furthermore, we would argue that the
photometric properties of the sample (Fig.~3) support our conclusion
that the reddest quasars in our sample are likely to be dust reddened
--- independent of our choice of definition.

Once we have isolated quasars that are probably dust reddened, it is
easy to divide the remainder of the objects into further color
classes.  Quasars with $\Delta(g^*-i^*)$ colors that are not in the
red tail of the distribution, $\Delta(g^*-i^*)\gtrsim0.3$, are
considered to be ``normal'' --- their range of colors is consistent
with that caused by a distribution of power-law continua.  We do not
impose a redshift dependent criterion to break the normal quasars into
color subclasses; we merely subdivide them into color classes along
lines of constant $\Delta(g^*-i^*)$.  See Table~\ref{tab:tab1} and
below for the definition of the four normal color quasar samples that
we will use in our analysis.

\section{Composite Spectra}

\subsection{Composite Spectra Construction \label{sec:compspec}}

We now create composite spectra as a function of relative color
(optical continuum slope) to study the dependence of spectral
properties on color in the ensemble average.  The composites are
constructed in the same fashion as the \markcite{vrb+01}{Vanden Berk} {et~al.} (2001) SDSS quasar
composite, using the same code.  In detail, the quasars are sorted by
their cataloged redshifts, shifted to their rest frame wavelengths,
rebinned to a common wavelength scale, scaled by the overlap of the
preceding average spectrum, and weighted by the inverse of the
variance.  The geometric mean of the spectra is used, which preserves
input power-law slopes [and $E(B-V)$ values, if the extinction curve
is the same in all objects].

Four samples of ``normal'' color quasars are created as a function of
color by separating all quasars with $\Delta(g^*-i^*) \leq 0.3$ into
quartiles in $\Delta(g^*-i^*)$; see Table~\ref{tab:tab1}.  Each
quartile contains 1053 quasars.  Next, we examine the redshift and
absolute magnitude distributions of each quartile in bins of $\Delta
M_i = 0.5$ and $\Delta z = 0.2$.  We restrict the number of quasars in
each redshift-magnitude bin to the smallest number in that bin for any
of the four quartiles.  Each quartile then has only 770 quasars, but
now the quartiles have roughly the same redshift and absolute
magnitude distribution, which allows us to compare differences between
the samples without having to worry that they are caused by redshift
or luminosity effects.  Table~\ref{tab:tab1} gives the number of
quasars in each of the subsamples, the $\Delta(g^*-i^*)$ color limits,
the spectral index of the composite spectra (see below), and the
number and percentage of FIRST-detected radio sources.

\begin{figure}[p]
\epsscale{1.0}
\plotone{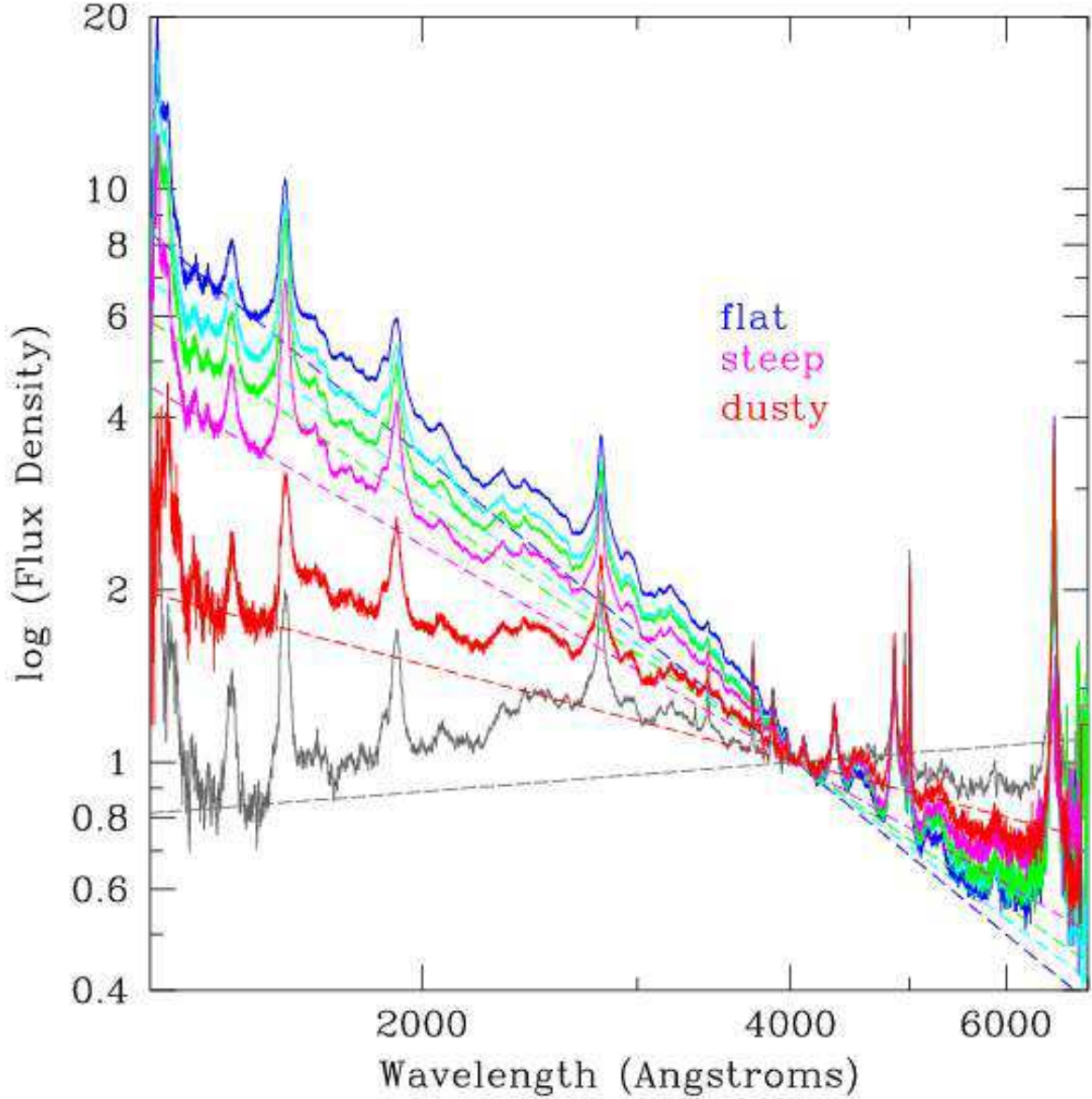}
\caption{Composite spectra as a function of $\Delta(g^*-i^*)$,
normalized at $4040\,{\rm \AA}$.  The blue, cyan, green, and magenta
spectra are the four normal quasar composite spectra that have similar
absolute magnitude and redshift distributions (composites 1-4).  The
red spectrum is the dust reddened composite (composite 5); it has a
different absolute magnitude and redshift distribution from the other
four spectra.  Composite 5 has significant curvature.  It is not well
fit by a power-law at $2200\,{\rm \AA}$ or for $\lambda \gtrsim
5000\,{\rm \AA}$.  The gray spectrum is a composite (number 6) of all
of the dust reddened quasars that were too red to be included in the
primary dust reddened composite; see text. \label{fig:fig7}}
\end{figure}

We create composite spectra for each of these four samples (hereafter
``composites 1-4''); these are shown in blue, cyan, green and magenta
in Figure~\ref{fig:fig7}.  The spectral indices (as measured from
$1450\,{\rm \AA}$ to $4040\,{\rm \AA}$) are $\alpha_{\nu} =$ $-0.25$,
$-0.41$, $-0.54$, and $-0.76$, respectively.  These values are less
reliable with increasing redness due to the presence of broad
absorption line troughs in the reddest samples (see the \ion{C}{4}
emission line regions in the upper left-hand panels of
Fig.~\ref{fig:fig8} and Fig.~\ref{fig:fig9}) and any possible
curvature induced by dust extinction.  That our photometrically
defined color samples yield composite spectra with power-law continua
in the same sense as their input colors testifies to the quality of
the SDSS's spectrophotometry.

We also create a composite dust reddened quasar spectrum (hereafter
``composite 5'') by combining all 211 spectra that meet the restricted
definition given above.  The dust reddened spectrum is shown in red in
Figure~\ref{fig:fig7}.  A single power-law is a poor fit to this
spectrum, as can be seen by examining the spectrum at
$\lambda\sim2200\,{\rm \AA}$ and $\lambda\gtrsim5000\,{\rm \AA}$.  In
addition, we present a composite spectrum of the 62 quasars that are
redder than the upper limit on the dust reddened composite that we
imposed above; this composite (hereafter ``composite 6'') is shown in
gray in Figure~\ref{fig:fig7}.  The quasars that contribute to the
dust reddened composite (composite 5) have a different distribution in
$M_{i}$ vs. $z$ space than the quasars in the normal color composites
that we defined above: on average, the spectra are $0.55$ mag fainter.
Thus, we have also created subsamples of the four normal quasar
samples and the dust reddened sample such that the four new normal
color samples and the new dust reddened sample all have the same
distribution in $M_{i}$ vs. $z$ space; see Table~\ref{tab:tab1}.
These five new composite spectra (hereafter ``composites 1n-5n'') have
185 quasars each and are similar to the original composites, but have
lower signal-to-noise, so we have not shown them in full (but see
Fig.~\ref{fig:fig9}).

\subsection{Composite Spectra Analysis} \label{sec:CSA}

\begin{figure}[p]
\epsscale{1.0}
\plotone{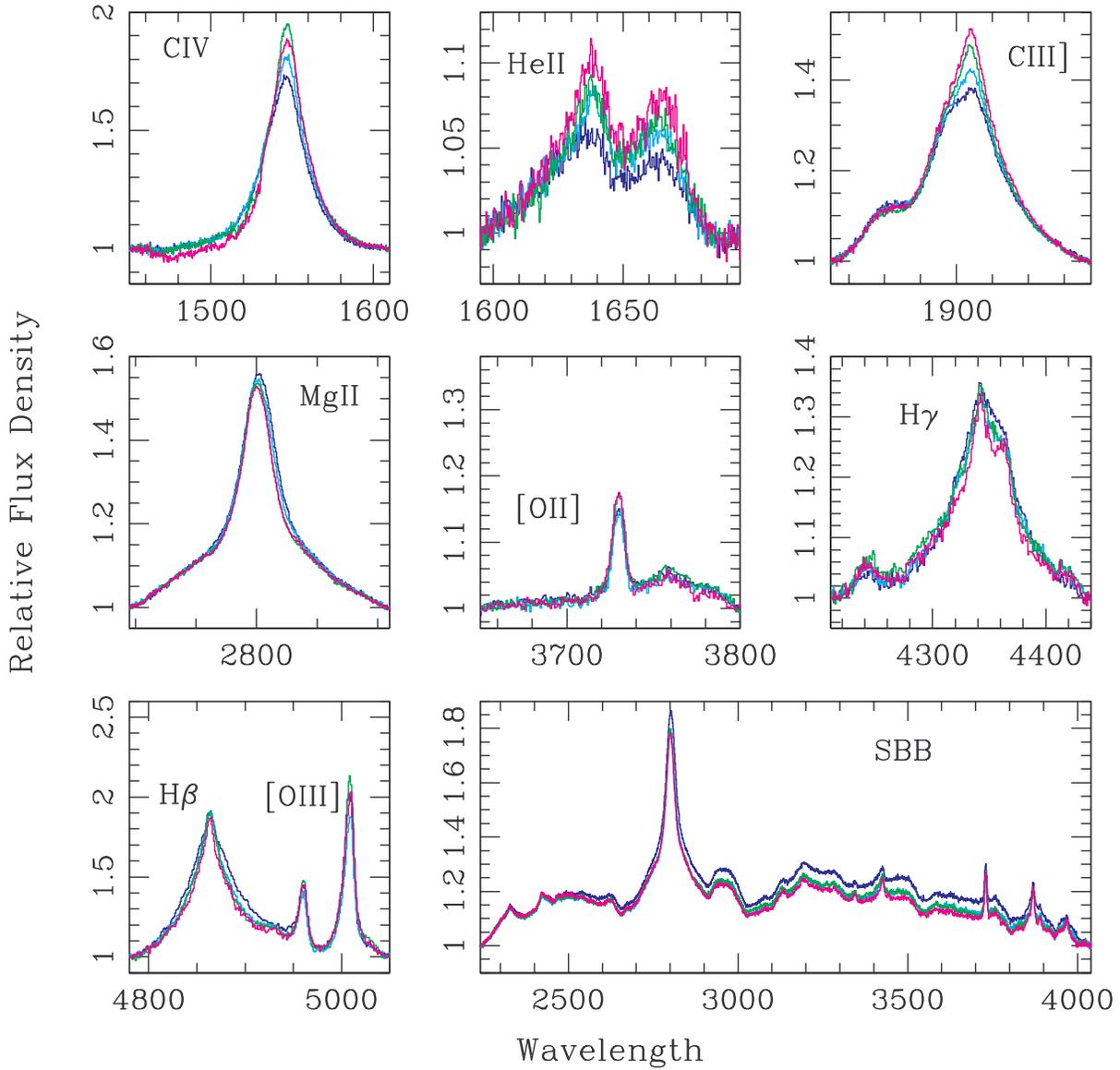}
\caption{Expanded emission line regions from Figure~\ref{fig:fig7}
(composites 1-4). ``SBB'' refers to the $3000\,{\rm \AA}$ (or small
blue) bump.  The color scheme is the same as for
Figure~\ref{fig:fig7}. The spectra are normalized such that their
continua agree at the edges of the panels. \label{fig:fig8}}
\end{figure}

\begin{figure}[p]
\epsscale{1.0}
\plotone{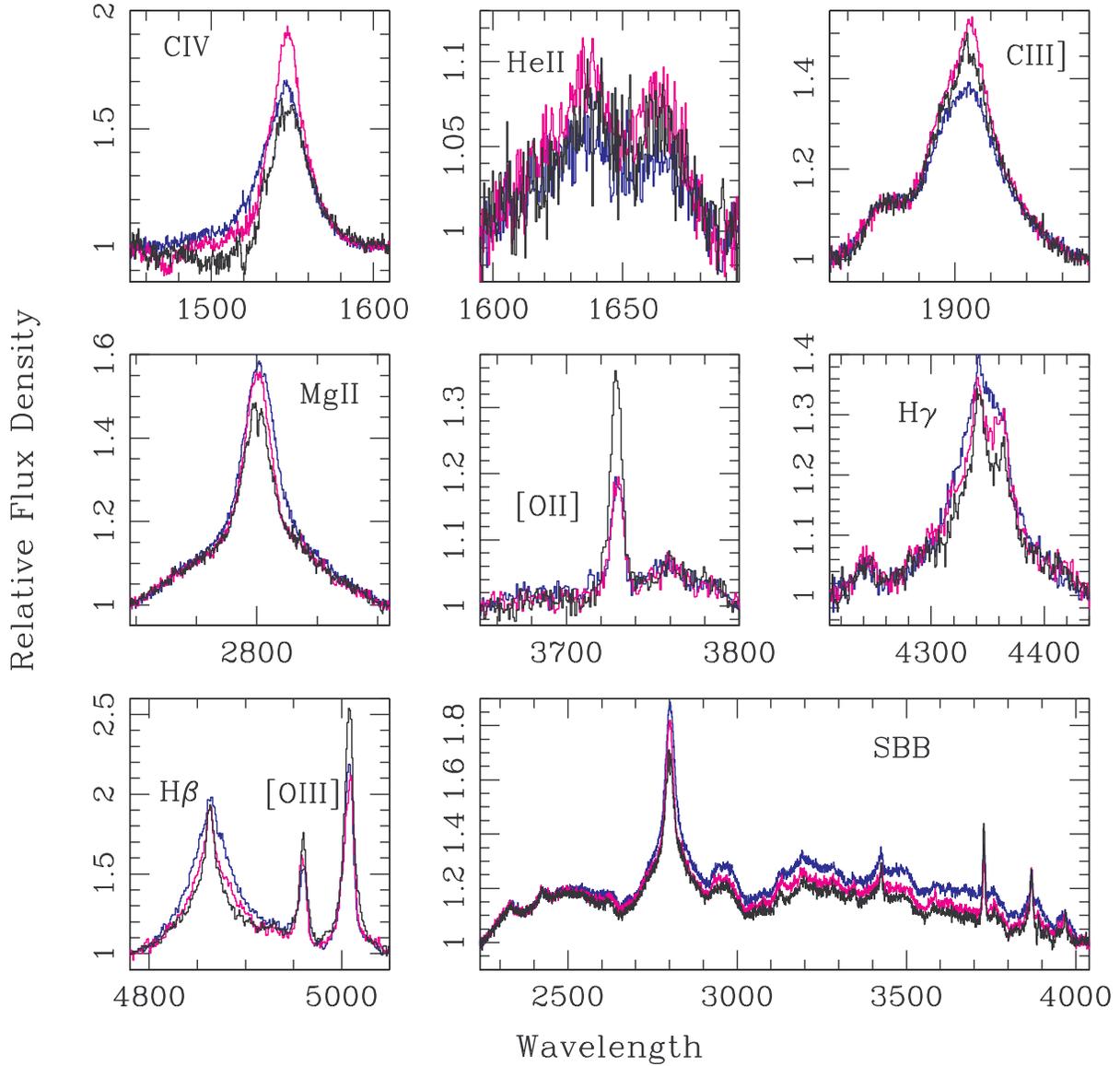}
\caption{Expanded emission line regions of the bluest and reddest
absolute magnitude and redshift normalized composite spectra
(composites 1n and 4n, cyan and magenta, respectively), and the
absolute magnitude and redshift normalized dust reddened composite
spectrum (composite 5n, black).  The spectra are normalized such that
their continua agree at the edges of the panels. \label{fig:fig9}}
\end{figure}

In addition to the change in the power-law spectral index with
$\Delta(g^*-i^*)$, some interesting changes occur to the emission line
features.  Here we point out some of the more obvious trends that will
be important for our later discussion.  The major emission line
regions of interest are shown in Figures~\ref{fig:fig8}
and~\ref{fig:fig9}.  These spectra have been normalized using local
continuum windows at the edges of the plots.  See Table~\ref{tab:tab2}
for the equivalent widths of the major emission lines in the color
composite spectra.  Below we comment on each of the panels in these
figures.  We begin by examining the four normal color composite
spectra (composites 1-4; Fig.~\ref{fig:fig8}) that are normalized to
have the same redshift and absolute magnitude
distribution.\footnote{As we will discuss further in the next section,
it is important to realize that any trends among the normal quasar
composites that are seen here will be distorted if quasars that are
intrinsically bluer than they appear are pushed into a redder sample
by dust reddening.}

\ion{C}{4} --- There is no clear systematic trend in \ion{C}{4} height
or width with color, although the redder spectra do seem to have
higher peaks and slightly wider profiles.  There {\em is}, however, a
clear trend in the existence of BAL-like absorption blueward of the
\ion{C}{4} emission line; the reddest quasars (composite 4) clearly
have more absorption on average, although the absorption feature is
diluted because of averaging with nonBALQSOs in the composite.  There
also appears to be a slight blueshift of the line (as seen in the red
wing of the line) with bluer colors; see also \markcite{rvr+02}{Richards} {et~al.} (2002b).

\ion{He}{2} --- The strength of \ion{He}{2}\,$\lambda1640\,{\rm \AA}$
grows with increasing redness.  The same is true for the
\ion{O}{3}]/\ion{Al}{2}/\ion{Fe}{2} blend just redward of the
\ion{He}{2} line.

\ion{C}{3}] --- There is a very clear systematic trend of increasing
\ion{C}{3}]\,$\lambda1909\,{\rm \AA}$ height with increasing redness.
The nearby \ion{Al}{3}\,$\lambda1857\,{\rm \AA}$ and
\ion{Si}{3}]\,$\lambda1892\,{\rm \AA}$ lines do not seem to show any
change with color.  The
\ion{Fe}{3}$\lambda\lambda$1895,1914,1926\,\AA\ complex also affects
this wavelength region, but the profile change cannot be explained
entirely by changes in the \ion{Fe}{3} strength since the four
composites agree well redward of the \ion{C}{3}] peak.

\ion{Mg}{2} --- \ion{Mg}{2} shows a weak, but clear trend towards
weaker lines with increasing redness.  There is also a slight
($\sim150\,{\rm km\,s^{-1}}$), but systematic shift of the line peak
to the blue with increasing redness.  Interestingly, this shift has a
color trend that is opposite to that of \ion{C}{4}.

[\ion{O}{2}] --- The two reddest composites have somewhat stronger
[\ion{O}{2}] $3727$ than the two bluest composites (see Table~2);
however, the differences are relatively small.

$H\gamma$ --- The $H\gamma$ line (which is blended with
[\ion{O}{3}]\,$\lambda$4363) is somewhat narrower in the reddest
composite spectrum.

$H\beta$ --- A systematic narrowing of the $H\beta$ line is seen with
increasing redness.  No systematic trend is seen for the [\ion{O}{3}]
lines.

$3000\,{\rm \AA}$ bump --- There exists a very clear trend towards a
weaker $3000\,{\rm \AA}$ (small blue) bump with increasing redness.
This trend is consistent with that seen by \markcite{bh95}{Baker} \& {Hunstead} (1995) for a sample
of radio-selected quasars, where both trends were a function of
increasing radio core-to-lobe flux density ratio $R$ --- that is,
orientation.  This issue is discussed further in the next section.

Unfortunately, we are not able to investigate $H\alpha$ (or the Balmer
decrement) since there are not enough quasars in each sample at this
wavelength.

The above trends are all with respect to the normal color composite
spectra.  Also of interest is how the dust reddened composite spectrum
fits into these trends.  Figure~\ref{fig:fig9} depicts the absolute
magnitude and redshift normalized dust reddened composite (5n) along
with composites 1n and 4n.  For these composites, we find the
following trends.  \ion{C}{4} is quite weak in the dust reddened
spectrum; both the emission line and the continuum just blueward of
the emission line are absorbed.  The \ion{C}{3}] emission line appears
intermediate in strength between the bluest and reddest composites.
The \ion{Mg}{2} emission line is weaker than the other color
composites and shows signs of intrinsic absorption with $z_{\rm abs}
\sim z_{\rm em}$.  [\ion{O}{2}] $3727$ is {\em much} stronger in the
dust reddened spectrum than in all of the other color composites.
Both $H\beta$ and $H\gamma$ are noticeably narrower in the dust
reddened composite spectrum.  [\ion{O}{3}] is considerably stronger in
the dust reddened spectrum.  The $3000\,{\rm \AA}$ bump in the dust
reddened spectrum is the weakest of all of the composites, continuing
the trend of a weaker feature with increased redness.

\section{Discussion}

\subsection{Red vs. Reddened \label{sec:redden}}

The shape of our putative dust reddened composite spectrum reveals
something fundamental about its nature.  The two primary causes of red
continua in the optical were discussed in detail by \markcite{fww00}{Francis} {et~al.} (2000).  In
their investigation of a sample of radio-detected quasars they showed
that if quasars are redder because of the addition of a synchrotron
turnover, then we expect them to be redder at red wavelengths than at
blue wavelengths (i.e., they will have ``u'' shaped continua).  On the
other hand, if the quasars are redder at blue wavelengths, then they
are reddened by dust extinction (i.e., they will have ``n'' shaped
continua).  Figures ~\ref{fig:fig3} and~\ref{fig:fig7} show that our
(mostly radio-quiet) quasars are are clearly redder at blue
wavelengths than at red wavelengths --- ``n" shaped --- and we
conclude that these red quasars are redder in a way that is consistent
with dust extinction and inconsistent with synchrotron emission.  We
have not fully investigated the nature of red radio-loud quasars
separately since they are few in our sample, but their color
distribution (see the two dashed histograms in Fig.~\ref{fig:fig3})
suggests that dust reddening is at work in radio-detected quasars as
well as radio-undetected ones.

We can then ask how red our dust reddened composite is relative to the
normal color composites.  Using an SMC-like reddening law with dust
local to the quasar, we find that the normal color composites (1n to
4n) need to be reddened by approximately $E(B-V)=0.135, 0.11, 0.10,
{\rm and}\, 0.07$, respectively from blue to red to fit the dust
reddened composite spectrum (5n).  A composite spectrum that includes
all of the quasars in samples 1n, 2n, 3n and 4n requires
$E(B-V)\simeq0.11$ to fit composite 5n.  These results are not
surprising given our dust reddened quasar definition, but the relative
values are still of interest.

\begin{figure}[p]
\epsscale{1.0}
\plotone{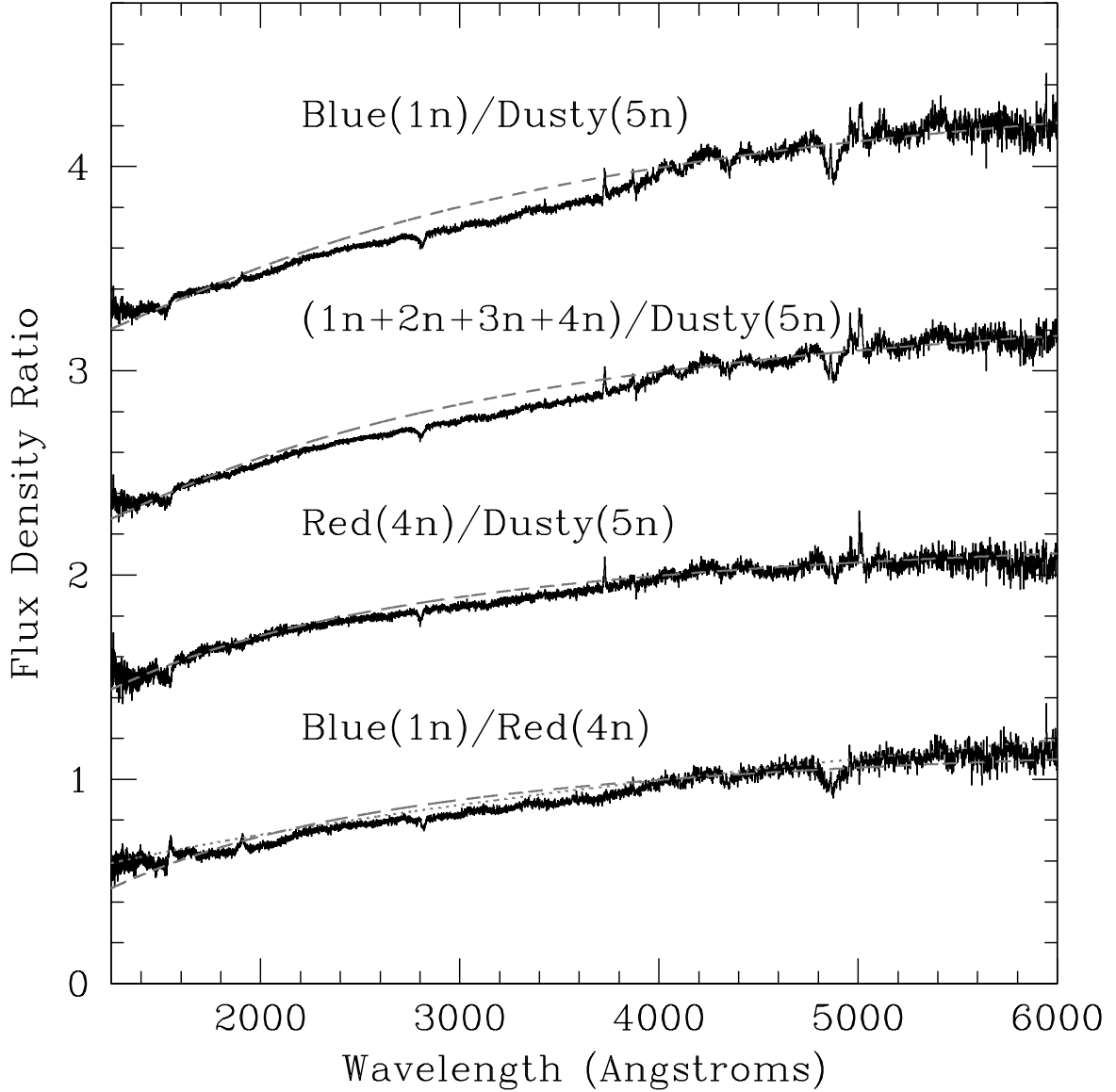}
\caption{Composite spectra ratios ({\em black lines}). The bottom
ratio spectrum is normalized to unity at $4040\,{\rm \AA}$; the top
three spectra are each shifted upwards from the preceding spectrum by
one unit in the flux density ratio.  For each ratio we overplot an SMC
reddening curve ({\em gray dashed lines}) with
$E(B-V)=0.135,0.11,0.07$, and $0.035$ from top to bottom,
respectively.  For the ratio of the bluest to reddest (optically
flattest and steepest; bottom panel) power-law composites we also
overplot a curve ({\em dotted gray line}) showing a change in spectral
index of $\Delta\alpha=0.45$. \label{fig:fig10}}
\end{figure}

Figure~\ref{fig:fig10} displays the ratios of composites 1n to 5n; the
average of 1n, 2n, 3n, and 4n to 5n; 4n to 5n; and 1n to 4n,
respectively from top to bottom.  Overplotted in gray dashed lines are
SMC reddening curves with $E(B-V)$ values of 0.135, 0.11, 0.07 and
0.035, respectively.  These reddening values result in good matches to
the dust reddened composite (5n) at both $1700\,{\rm \AA}$ and
$4040\,{\rm \AA}$, but they overpredict the flux between these
wavelengths and underpredict the flux shortward of \ion{C}{4}.  This
discrepancy between these wavelengths is most likely the result of
differences in the strength of the Balmer continuum between the
composites.  If we modeled these Balmer continuum differences, the
resulting SMC reddening curves would trace the overall continuum
differences much more closely.  Similar results are found when using a
simple $1/\lambda$ reddening law as given by \markcite{fww00}{Francis} {et~al.} (2000).  Reddening
the normal color composites with the shallower LMC reddening curve
produces only marginally better fits to the dust reddened composite
spectrum.

The issue of the color differences between the normal composites is
somewhat more complicated.  To help address this question, in the
bottom panel of Figure~\ref{fig:fig10}, we overplot a curve ({\em
dotted gray line}) showing a difference in spectral index of
$\Delta\alpha=0.45$ on the ratio of the bluest to the reddest of the
power-law composites.  Comparing this curve to the curve produced by
pure SMC reddening ({\em dashed gray line}) suggests that it is likely
that dust reddening plays {\em some} role in the color differences
between composites 1-4, since the ratio spectrum has slightly more
curvature than the spectral index difference curve.  Indeed, we will
argue in \S~\ref{sec:trends} that the \ion{C}{4} emission line shape
as a function of color suggests that some intrinsically blue
(optically flat) quasars are being reddened by dust into the redder
(steeper) color bins.  At the same time, reddening alone cannot be the
cause of the significant trends in emission line properties from
composite 1 to composite 4.  Either the emission lines are responding
directly to differences in the ionizing continuum, or both the optical
SED and the emission lines are changing as a function of some other
property (such as orientation or accretion rate).  As a result, we
suggest that the color changes between composites 1 through 4 are
dominated by changes in the intrinsic continuum rather than by dust
reddening (see \S~\ref{sec:trends}).

Finally, these ratio spectra in Figure~\ref{fig:fig10} also highlight
some of the changes in the emission line regions that we discussed
earlier.  One new result from this presentation is that the narrow
component of H$\beta$ seems to be relatively constant while the broad
component is changing (as can be seen by the inflection in the H$\beta$
residual profile).

\subsection{BALQSOs and Reddening \label{sec:balred}}

It has been noted by numerous authors that BALQSOs, especially LoBALs,
are redder than the average quasar
\markcite{sf92,btv+97,yv99,bwg+00,ndb+00,btb+01,hal+02,rei+02}({Sprayberry} \& {Foltz} 1992;  ; {Yamamoto} \& {Vansevi{\v c}ius} 1999; {Becker} {et~al.} 2000; {Najita}, {Dey}, \& {Brotherton} 2000; {Brotherton} {et~al.} 2001; {Hall} {et~al.} 2002a; {Reichard et al.} 2003), likely as
a result of extinction by dust.  The reddest of our four color
composites and the dust reddened composite spectrum show considerable
absorption blueward of \ion{C}{4}, in agreement with this result.

An interesting issue is the BALQSO fraction as a function of optical
color; we explore this question using a well-defined BALQSO catalog
\markcite{rrh+02}({Reichard} {et~al.} 2003) from the SDSS Early Data Release (EDR;
\markcite{slb+02}{Stoughton} {et~al.} 2002).  First, we consider only those quasars that would
have been selected by the \markcite{rfn+02}{Richards} {et~al.} (2002a) algorithm since the EDR
quasar sample was not selected with one uniform algorithm.  Second, we
consider only quasars with $1.7 \le z \le 2.2$ so that the sample is
restricted to those quasars for which a BALQSO classification was made
using spectra of the \ion{C}{4} absorption region.  Finally, we divide
the quasars into four equal bins of normal color quasars (with 434
quasars in each bin) in the same manner as for our primary sample, and
we also define an EDR dust reddened sample with 96 quasars.  We find
that from bluest to reddest (optically flattest to steepest), the
BALQSO fractions are $3.4\pm1.5$\%, $6.3\pm2.3$\%, $10.3\pm2.7$\%, and
$18.1\pm3.5$\%.  The BALQSO fraction among EDR dust reddened quasars
is is $20.0\pm8.9$\%, after restricting the samples such that they
have the same absolute magnitude and redshift distributions.

Thus, in agreement with the work cited above, we find that the
presence of BAL troughs and reddening are highly correlated.  It is
unclear to us whether the BAL material always reddens the intrinsic
quasar spectrum, or if reddened quasars are more likely to be BALQSOs,
but there is little doubt that BALQSOs are indeed reddened in a manner
that is consistent with dust extinction.  If BALQSOs suffer from dust
extinction, then the fractions as a function of color given above are
not the {\em intrinsic} fractions.  That is, BAL troughs may be just
as common in intrinsically blue (optically flat) quasars as in
intrinsically red (optically steep) quasars if dust extinction skews
the observed distribution of colors to the red.

In this analysis, we also must be careful about how any broad
absorption troughs affect the colors of BALQSOs since the \ion{C}{4}
trough will often fall within the $g$ band.  For the majority of
BALQSOs the change in color due to reddening in the $g$-band will
dominate over the change in color due to an absorption trough in the
$g$-band, as can be seen from the following argument.  The standard
measure of \ion{C}{4} BAL strength is the ``balnicity'' index, which
is sensitive to flux decrements of 10\% or more over a span of
$25,000\,{\rm km\,s^{-1}}$ (from the peak of \ion{Si}{4} to the peak
of \ion{C}{4}, or roughly $450\,{\rm \AA}$ at $z=2$) \markcite{wmf+91}({Weymann} {et~al.} 1991).
On the other hand, a $z=2$ quasar with $E(B-V)$ of only $0.03$ has a
flux decrement across the {\em entire} $g$-band of 14\% (relative to
the $i$-band, and assuming SMC-like intrinsic reddening).  Since the
$g$-band has an effective width that is more than twice the wavelength
range considered when calculating balnicities and since BAL troughs
rarely span as much as $25,000\,{\rm km\,s^{-1}}$ \markcite{rrh+02}({Reichard} {et~al.} 2003), the
effective absorption due to reddening completely dominates that caused
by absorption troughs.  Our use of relative $g^*-i^*$ colors as a
tracer of continuum color thus is valid for the vast majority of
BALQSOs, failing only for FeLoBALs and for LoBALs with very extensive
absorption \markcite{hal+02}({Hall} {et~al.} 2002a).

\subsection{Emission Line Trends with Color \label{sec:trends}}

Many of the emission line trends with color discussed in
\S~\ref{sec:CSA} can be related to the principal component analysis
(PCA) of quasars, which has become popular in recent years
\markcite{bg92,smz+02}({Boroson} \& {Green} 1992; {Sulentic} {et~al.} 2002).  The basic idea of PCA is to describe the data by
orthogonal eigenvectors and to study the resulting distribution of
eigenvalues to learn something about the underlying physics of
quasars.  Of particular interest are the correlations with the first
eigenvector as found by \markcite{bg92}{Boroson} \& {Green} (1992) and summarized by \markcite{bf99}{Brotherton} \& {Francis} (1999).

One of the key emission lines used in eigenvector analysis is
[\ion{O}{3}] (see \markcite{bf99}{Brotherton} \& {Francis} 1999, and references therein).  In our
sample, little difference is seen in [\ion{O}{2}] and [\ion{O}{3}]
among the normal color composites; however, both of these lines are
significantly stronger in the dust reddened composite.  This result
suggests that optically steep quasars are not produced exclusively by
dust reddening, but that the dust reddened spectrum is affected since
the increase in line strength is larger at
[\ion{O}{2}]\,$\lambda3727\,{\rm \AA}$ than at
[\ion{O}{3}]\,$\lambda5007\,{\rm \AA}$ --- as expected in a simple
dust reddening scenario where dust affects the continuum and broad
emission line region (BELR), but not the narrow emission line region
\footnote{An even simpler scenario where the dust affects the narrow
line region as well is ruled out because it predicts identical
equivalent widths for the narrow lines in all composites.}.

We can test the consistency of this scenario by comparing the
[\ion{O}{2}] and [\ion{O}{3}] EWs predicted for a reddened normal
quasar with the observed EWs of the dust reddened composite.  The
[\ion{O}{3}]$\lambda$5007 EW for the dust reddened quasars is
18.97\,\AA, while the average for normal quasars (composites 1n-4n) is
14.15\,\AA.  Assuming the continuum is extincted but the narrow-line
region is not, to match this difference requires $E(B-V)=0.10$ with
the SMC extinction curve, in good agreement with the extinction
required to best match the normal quasar composite to the dust
reddened composite.  However, the [\ion{O}{2}]$\lambda$3727 EW for
dust reddened quasars is 2.81\,\AA\ (after correcting for the 5\%
weaker Balmer decrement in the dust reddened composite at 3727\,\AA),
while the average for normal quasars is 1.39\,\AA, and to match this
difference requires $E(B-V)=0.17$.

This discrepancy could be explained if dust reddened quasars have
stronger [\ion{O}{2}] emission than normal quasars.  \markcite{crc+02}{Croom} {et~al.} (2002)
suggest that a significant portion of the [\ion{O}{2}] flux in quasars
arises from the host galaxy.  If true, stronger [\ion{O}{2}] in dust
reddened quasars might be expected, as galaxies with significant dust
content are also likely to have significant star formation.  Another
possibility is that [\ion{O}{3}] emission is less isotropic (or more
anisotropic) than that of [\ion{O}{2}].  Such anisotropy could occur
if the higher-ionization [\ion{O}{3}]-emitting region extended to
small enough radii to be partially obscured by the dust which reddens
the quasars (e.g., a dusty torus viewed at high inclination).  This
latter geometry has been suggested for radio-loud quasars
\markcite{hbf96}({Hes}, {Barthel}, \& {Fosbury} 1996) though it may not apply to radio-quiet quasars
\markcite{kwb+00}({Kuraszkiewicz} {et~al.} 2000).  Note that in this latter case there would have to be
a range of reddenings across the sightline(s) to the continuum source
\markcite{hw95}({Hines} \& {Wills} 1995) to explain why the continuum appears reddened by only
$E(B-V)=0.1$, instead of $E(B-V)=0.17$.

\markcite{bg92}{Boroson} \& {Green} (1992) also noted that there is an anti-correlation between the
strengths of [\ion{O}{3}] and \ion{Fe}{2}.  If the small blue bump is
dominated by \ion{Fe}{2} emission, the small blue bump being weak in
dust reddened quasars when [\ion{O}{3}] is strong would be consistent
with this anti-correlation.  However, as can be seen in the ratio
spectra in Figure~\ref{fig:fig10}, the difference between composite 1n
and the dust reddened composite (5n) has a minimum near
$3500-3700\,{\rm \AA}$, rises steadily towards the blue, and rises
more rapidly to the red.  This behavior is consistent with this
color-dependent component of the small blue bump being dominated by
the Balmer continuum as opposed to \ion{Fe}{2} and \ion{Fe}{3}
emission \markcite{wnw85}({Wills}, {Netzer}, \& {Wills} 1985).

Although \markcite{bg92}{Boroson} \& {Green} (1992) argue that their ``Eigenvector 1'' is
independent of orientation, it is interesting to examine these results
from an orientation perspective by comparing our results to the
results of investigations of radio-loud quasars.  \markcite{bh95}{Baker} \& {Hunstead} (1995) and
\markcite{bak97}{Baker} (1997) argue that radio-loud quasars (RLQs) of increasing lobe
dominance are seen through increasing amounts of dust (i.e., RLQs are
redder the closer they are to edge-on) and are observed to have weaker
small blue bumps.  The weaker small blue bump, stronger oxygen lines,
and greater EW increase in [\ion{O}{2}] than in [\ion{O}{3}] seen in
our dust reddened composite could be consistent with their argument
(but see \markcite{kwb+00}{Kuraszkiewicz} {et~al.} 2000).  It has also been suggested that an
orientation effect would result if $H\beta$ were emitted from a
disk-like structure \markcite{wb86,vwb00}({Wills} \& {Browne} 1986; {Vestergaard}, {Wilkes}, \&  {Barthel} 2000) with broader lines coming from
quasars seen more edge-on.  Curiously, however, if the change in the
width of the Balmer lines is a function of orientation, then our
results --- narrower lines in objects which are presumably closer to
edge-on (optically steep and dust reddened quasars) --- conflict with
\markcite{wb86}{Wills} \& {Browne} (1986).  These conflicting results may indicate a real
difference between the BELRs of radio-loud and radio-quiet quasars.
Another possible explanation for the change in width of H$\beta$ with
color is a correlation with black hole mass \markcite{ves02}({Vestergaard} 2002); perhaps the
bluer (flatter) quasars simply have more massive black holes.

In an investigation of the X-ray properties of quasars, \markcite{lfe+97}{Laor} {et~al.} (1997)
found a correlation between the width of H$\beta$ and the X-ray
spectral index in the sense that narrow H$\beta$ corresponds to a soft
X-ray dominated spectrum and higher $L/L_{Edd}$ (though it is unclear
whether the X-ray SED is due to a soft X-ray excess or an
intrinsically steep X-ray spectrum).  Given our own results, this
finding implies that redder quasars may have steeper X-ray spectral
indices or larger soft X-ray excesses.  Similarly, we have weak
evidence that redder quasars are more likely to be radio sources than
the bluer quasars (see Table~\ref{tab:tab1}).  This result is
consistent with \markcite{rvr+02}{Richards} {et~al.} (2002b) if redder quasars also have smaller
\ion{C}{4} blueshifts (which will require a larger sample to
determine).  On the other hand, \markcite{smz+02}{Sulentic} {et~al.} (2002) found that it was the
radio-quiet quasars that tended to have the narrowest H$\beta$
emission lines and the largest soft X-ray excesses.  Clearly more work
is needed to understand the X-ray and radio properties as a function
of color.

Another particularly interesting dilemma with regard to the emission
lines as a function of color involves \ion{C}{4}, \ion{He}{2} and
\ion{C}{3}], all of which are popular lines in eigenvector analyses.
\markcite{rvr+02}{Richards} {et~al.} (2002b) found that quasars with large \ion{C}{4} blueshifts
with respect to \ion{Mg}{2} tended to have weaker \ion{C}{4},
\ion{He}{2} and \ion{C}{3}] emission and also tended to be bluer than
quasars with small \ion{C}{4} blueshifts.  Similarly, this paper's
color-divided samples also show that the bluest quasars have the
weakest \ion{C}{4}, \ion{He}{2}, and \ion{C}{3}] emission.
\markcite{rvr+02}{Richards} {et~al.} (2002b) argued that the emission line properties of such
quasars were most similar to BALQSOs (or vice versa), whereas in our
present work, it is the reddest quasars that are most likely to be
BALQSOs.  This trend with color for these three emission lines does
not seem to extend to the dust reddened sample, which contains the
most BALQSOs (see \S~\ref{sec:balred}).  This discrepancy may be
reconciled if dust reddened, but intrinsically blue quasars
contaminate the redder composites, including the dust reddened
composite, and if BALQSOs are drawn from a bluer parent population
\markcite{rei+02}({Reichard et al.} 2003).

\subsection{The Fraction of Missing Quasars}

We have not formally addressed the question of what fraction of
quasars are missed by normal optically-selected samples of quasars
because of reddening or because they would be classified as ``Type 2''
(narrow line only) quasars \markcite{nhg+02}(e.g., {Norman} {et~al.} 2002).  Estimates for the
fraction of optically extincted quasars that are ``missing'' have
ranged as high as 80\% for radio-quiet quasars, based on the existence
of red radio-loud quasars \markcite{wfp+95}({Webster} {et~al.} 1995).  In addition, X-ray surveys
estimate that a significant fraction of AGNs are undetected in the
optical if obscured AGNs (particularly Type 2 quasars) are the primary
contributors to the X-ray background \markcite{mcb+00,bhs+00}({Mushotzky} {et~al.} 2000; {Brandt} {et~al.} 2000).

Our red quasar sample can be used to reveal something about the
population of reddened quasars that might be missed by blue-sensitive
flux-limited surveys, but that are recovered by the SDSS.  A full
characterization of the incompleteness of previous UVX quasar surveys
to dust reddened quasars is beyond the scope of this paper.  We merely
point out that the SDSS {\em must}, by definition, be more sensitive
to dust reddened quasars simply because the SDSS is flux limited in
the $i$-band instead of the $B$-band and because dust extinction is
stronger at bluer wavelengths than red.  Yet even our sample must be
incomplete since despite the SDSS's sensitivity to red quasars and its
$i$-band selection, most quasars that are heavily dust reddened
[$E(B-V)\gtrsim0.5$] will be extincted below the limiting flux of
even the SDSS.

Furthermore, our sample is restricted to those objects that were
automatically classified as quasars by the SDSS's spectroscopic
pipeline.  Quasars that have absent, weak, or very narrow emission
lines; or that are extremely reddened, or show strong intrinsic
absorption, such as those described by \markcite{hal+02}{Hall} {et~al.} (2002a), are missing
from this sample.  Examples of extremely reddened quasars are rare,
but they carry considerable weight since they are intrinsically
luminous and are the only representatives in the SDSS of the
population of similarly reddened but less luminous quasars.  Thus,
they must also be considered when determining the fraction of missing
quasars.  A full analysis of this issue is beyond the scope of this
paper (and data set), but with the data in hand, we can perform some
preliminary incompleteness calculations.

We can make a rough estimate of the fraction of missing quasars with
$E(B-V)$ values within the limits of detectability by the SDSS.  We
take $E(B-V)$=0.11 (see \S 5.1) as representative of the dust reddened
quasar population relative to normal quasars.  At a representative
$z=2$, observed $g$ and $i$ correspond to roughly 1550\,\AA\ and
2500\,\AA\ in the rest frame.  An SMC extinction law with
$E(B-V)$=0.11 produces $A_{1550}$=1.43 and $A_{2500}$=0.81 and thus
reddens the observed $g^*-i^*$ color by $\Delta(g^*-i^*)=0.62$.  At a
given apparent magnitude, for this representative SDSS-detected dust
reddened quasar we are sampling the $i$-band luminosity function 0.8
magnitudes brighter than for normal quasars.  In the luminosity range
of our quasar sample, the 2dF QSO luminosity function \markcite{bsc+00}({Boyle} {et~al.} 2000)
shows an increase in the number of quasars of about a factor of 3 per
unit magnitude.  We must thus correct the number of dust reddened
quasars\footnote{In this analysis, we do not impose the upper limit to
the redness of the dust reddened quasars that we imposed when making
the composite spectrum.} with $\Delta(g^*-i^*)=0.6$ by a factor of
2.5, $\Delta(g^*-i^*)=1.2$ by a factor of 6.25, and so on.  There are
only 9 dust reddened quasars with $\Delta(g^*-i^*)>0.9$ in our sample
which require this factor 6.25 correction (Fig.~\ref{fig:fig6}); the
remaining 264 dust reddened quasars will have a factor 2.5 correction.
Thus, in addition to the 273 dust reddened quasars detected in our
SDSS sample, constituting 6\% of that sample (4.5\% if we restrict the
sample to $i^*<19.1$), we estimate that there are a further 443 (a
further 10\%) which would have been included in our SDSS sample if
they had not been reddened.  This fraction, however, assumes that our
quasars have magnitudes near our flux limit and therefore somewhat
overestimates the actual fraction of dust reddened quasars that the
SDSS might miss.  Nevertheless, this estimate, while crude,
illustrates the importance of accounting for the reddening of such
quasars when studying the full extent of the quasar population.

We can also make more general conclusions about the total AGN
population.  Specifically, the distribution of colors in our optical
data suggests that the hidden X-ray population cannot be due to a
continuous distribution of dust reddened, but otherwise normal Type 1
quasars.  If the fraction of missing dust reddened Type 1 quasars were
near 50\%, the $\Delta(r^*-i^*)$ and $\Delta(i^*-z^*)$ color
distributions shown in Figure~\ref{fig:fig3} would be skewed to the
red.  Thus, if the X-ray background is due to obscured AGN, there
would either have to be a class of broad-lined quasars whose reddening
distribution is disjoint from the distribution that we observe, or the
quasars would have to have their broad-line region obscured and thus
would be classified as Type 2 quasars (as is generally
suspected)\footnote{For a different conclusion, from a much smaller
sample, see \markcite{whb+03}{White et al.} (2003).}.  In fact, another way of viewing the
problem is described by \markcite{imk+02}{Ivezi{\' c}} {et~al.} (2002) who used FIRST-SDSS matches to
show that the SDSS quasar survey is at least 89\% complete to
broad-lined AGN.

\section{Conclusions}

We have presented a homogeneously selected sample of 4576 SDSS quasars
with $0.3 \le z \le 2.2$ and $i^*<19.9$.  The majority of these
quasars are well-fit by a power-law continuum with spectral indices
ranging from $\alpha_{\nu}=-0.25$ to $\alpha_{\nu}=-0.76$ (optically
flat to optically steep).  There is also a population of quasars that
are red in comparison with even the reddest power-law continuum
(optically steep) quasars and that are probably dust reddened.  This
population is $\simeq6.0\%$ of the sample, a large fraction of which
are likely to be missed by blue-sensitive optical surveys for quasars.
A rough correction for incompleteness due to the extinction in this
population suggests that the true fraction of dust reddened red
quasars to which the SDSS quasar survey would otherwise be sensitive
is $\simeq15\%$.

Among the power-law continuum quasars, optically steep quasars are
observed to have stronger emission at \ion{C}{4}, \ion{C}{3}], and
\ion{He}{2}.  The strength of the $3000\,{\rm \AA}$ bump is weaker in
optically steep quasars than in optically flat quasars.  This change
in strength appears to be dominated by weakening of Balmer emission
rather than iron emission.  In addition to weaker Balmer continuum,
redder quasars (including the dust reddened ones) have narrower Balmer
lines.  For the dust reddened quasar sample, we also find that
[\ion{O}{2}] and [\ion{O}{3}] emission have very large equivalent
widths.

We also find that BALQSOs appear to be more common among red quasars
than among blue quasars and that this color distribution cannot be
caused by the absorption troughs themselves.  The BALQSO fraction is
as high as 20\% among the dust reddened quasars and as low as 3.4\% in
optically flat quasars.  However, these fractions fail to account for
the possibility that BALQSOs may be systematically reddened by dust
and may have an intrinsically bluer parent population.

Further work on both the color distribution of quasars and the
population of dust reddened quasars will benefit considerably from the
availability of much larger samples of SDSS quasars in the near
future, and from the extension of our technique to somewhat higher
redshifts.  In particular, we will be able to study the very reddest
quasars found by the SDSS, which are much redder than those considered
herein.  We will also be better able to explore the questions of
whether the reddening is always due to dust, how often the dust is
intervening rather than intrinsic, and what the form of the average
reddening law is in both cases.

\acknowledgements

Funding for the creation and distribution of the SDSS Archive has been
provided by the Alfred P. Sloan Foundation, the Participating
Institutions, the National Aeronautics and Space Administration, the
National Science Foundation, the U.S. Department of Energy, the
Japanese Monbukagakusho, and the Max Planck Society. The SDSS Web site
is http://www.sdss.org/.  The SDSS is managed by the Astrophysical
Research Consortium (ARC) for the Participating Institutions. The
Participating Institutions are The University of Chicago, Fermilab,
the Institute for Advanced Study, the Japan Participation Group, The
Johns Hopkins University, Los Alamos National Laboratory, the
Max-Planck-Institute for Astronomy (MPIA), the Max-Planck-Institute
for Astrophysics (MPA), New Mexico State University, University of
Pittsburgh, Princeton University, the United States Naval Observatory,
and the University of Washington.  This work was supported in part by
National Science Foundation grants AST99-00703 (G.~T.~R. and D.~P.~S.)
and AST00-71091 (M.~A.~S.).  P.~B.~H. is supported by FONDECYT grant
1010981.  We thank Tim Heckman, Julian Krolik, and Bob Becker for
discussions that contributed to this paper.  We also thank an
anonymous referee for suggestions that led to improvements in the
presentation of the paper.

\clearpage



\clearpage

\clearpage

\begin{deluxetable}{lcllll}
\tabletypesize{\small}
\tablewidth{0pt}
\tablecaption{Sample Definitions \label{tab:tab1}}
\tablehead{
\colhead{Sample} &
\colhead{${\rm N_{QSOs}}$} &
\colhead{$\Delta(g^*-i^*)_{\rm min}$} &
\colhead{$\Delta(g^*-i^*)_{\rm max}$} &
\colhead{$\alpha$} &
\colhead{Radio\tablenotemark{a}}
}
\startdata
1 & 770 & $-$0.461 & $-$0.109 & $-$0.25 & 18/3.0\\
2 & 770 & $-$0.109 & $-$0.032 & $-$0.41 & 32/5.5\\
3 & 770 & $-$0.031 & +0.066 & $-$0.54 & 26/4.4\\
4 & 770 & +0.066 & +0.299 & $-$0.76 & 51/8.7\\
5 & 211 & $+0.200$ & $+0.200$ & \nodata & \nodata\\
& & $ +E(B-V)\ge0.04$ & $ +E(B-V)\le0.12$ & & \\
6 & 62 & $+0.200$ & \nodata & \nodata & \nodata\\
\smallskip
& & $ +E(B-V)\ge0.12$ & \nodata & & \\
1n & 185 & $-$0.461 & $-$0.109 & \nodata & 0/0.0\\
2n & 185 & $-$0.109 & $-$0.032 & \nodata & 7/4.8\\
3n & 185 & $-$0.031 & +0.066 & \nodata & 6/3.9\\
4n & 185 & +0.066 & +0.299 & \nodata & 9/6.0\\
5n & 185 & $+0.200$ & $+0.200$ & \nodata & 16/10.2\\
 & & $ +E(B-V)\ge0.04$ & $ +E(B-V)\le0.12$ & & \\
\enddata
\tablenotetext{a}{Number and percentage of FIRST-detected radio sources.  Not all of the quasars lie in an area covered by FIRST.}
\end{deluxetable}

\begin{deluxetable}{lrrrrrrrrrr}
\tabletypesize{\small}
\tablewidth{0pt}
\tablecaption{Color Composite Emission Line Equivalent Widths \label{tab:tab2}}
\tablehead{
\colhead{} &
\colhead{Si~IV}&\colhead{C~IV}&\colhead{He~II}&\colhead{C~III]}&\colhead{Mg~II}&
\colhead{[O~II]}&\colhead{H$\gamma$}&\colhead{H$\beta$}&\colhead{[O~III]}&\colhead{[O~III]}\\[.2ex]
\colhead{Spectrum} &
\colhead{1400}&\colhead{1550}&\colhead{1640}&\colhead{1909}&\colhead{2800}&
\colhead{3727}&\colhead{4340}&\colhead{4862}&\colhead{4959}&\colhead{5007}
}
\startdata
1 & 8.44 & 23.80 & 0.46 & 19.12 & 32.48 & 0.88 & 19.15 & 37.82 & 3.34 & 11.57 \\
2 & 8.20 & 25.80 & 0.44 & 20.09 & 31.61 & 0.97 & 18.34 & 34.64 & 3.11 & 11.31 \\
3 & 8.15 & 27.36 & 0.57 & 20.43 & 30.14 & 1.26 & 15.47 & 33.52 & 3.68 & 15.51 \\
4 & 9.15 & 26.86 & 0.91 & 21.45 & 30.44 & 1.30 & 14.99 & 29.93 & 3.70 & 13.43 \\
\smallskip
5 & 9.19 & 19.03 & 0.54 & 21.49 & 28.97 & 2.95 & 15.50 & 26.15 & 5.52 & 18.97 \\
1n & 6.85 & 24.00 & 0.21 & 17.76 & 33.11 & 1.54 & 19.86 & 39.73 & 4.42 & 15.30 \\
2n & 6.57 & 24.64 & 0.63 & 20.51 & 31.73 & 1.20 & 17.99 & 37.82 & 3.27 & 12.98 \\
3n & 9.58 & 24.87 & 0.43 & 18.64 & 31.43 & 1.46 & 18.59 & 34.84 & 3.67 & 14.16 \\
4n & 8.75 & 25.50 & 0.67 & 23.37 & 30.62 & 1.37 & 15.05 & 29.23 & 5.09 & 14.17 \\
\enddata
\tablecomments{The wavelengths and equivalent widths are given in
Angstroms and are measured in the rest frame.}
\end{deluxetable}

\end{document}